\documentclass[fleqn,usenatbib]{mnras}

\usepackage{subfig}

\usepackage{amsfonts,amssymb,amsbsy,amsmath}
\usepackage{mathtools}
\usepackage{upgreek}

\usepackage{natbib}
\usepackage{graphicx}
\usepackage{xifthen}
\usepackage{bigints}
\usepackage{booktabs}
\usepackage{nicefrac}
\usepackage{color}
\usepackage{url}
\usepackage{hyperref} 
\usepackage[theorems]{tcolorbox}

\catcode`_=\active
\newcommand_[1]{\ensuremath{\sb{\mathrm{#1}}}}

\newcommand{\newpar}{{\\}}

\newcommand{\bs}{\boldsymbol}

\newcommand{\numChar}{{n}} 

\newcommand{\paraChar}{{p}}

\newcommand{\episChar}{{n}}

\newcommand{\npe}[1][]{{ \numChar_{ \ifthenelse{\isempty{#1}}{\paraChar\episChar}{{\paraChar\episChar,#1}} } }} 

\newcommand{\epis}{{nois}}

\def\gtrsim{\mathrel{\hbox{\rlap{\hbox{\lower4pt\hbox{$\sim$}}}\hbox{$>$}}}}
\def\lessim{\mathrel{\hbox{\rlap{\hbox{\lower4pt\hbox{$\sim$}}}\hbox{$<$}}}}

\newcommand{\rmz}{{\rm z}}

\newcommand{\emodel}[1][]{{ M_{ \ifthenelse{\isempty{#1}}{\epis}{{\epis,#1}} } }}
\newcommand{\emodelz}[1][]{{ M^\rmz_{ \ifthenelse{\isempty{#1}}{\epis}{{\epis,#1}} } }}
\newcommand{\emodellgrb}[1][]{{ M^\lgrb_{ \ifthenelse{\isempty{#1}}{\epis}{{\epis,#1}} } }}

\newcommand{\lgrb}{{\rm LGRB}}

\newcommand{\param}{{\bs{\theta}}}

\newcommand{\eparam}[1][]{{ \param_{ \ifthenelse{\isempty{#1}}{\epis}{{\epis,#1}} } }}
\newcommand{\eparamz}[1][]{{ \bs\param^\rmz_{ \ifthenelse{\isempty{#1}}{\epis^\rmz}{{\epis,#1}} } }}
\newcommand{\eparamlgrb}[1][]{{ \bs\param^\lgrb_{ \ifthenelse{\isempty{#1}}{\epis^\lgrb}{{\epis,#1}} } }}

\newcommand{\truth}{{\bs{R}}}
\newcommand{\possible}{{*}}
\newcommand{\truthset}{{\mathcal{R}}}

\newcommand{\truthsubset}[1][]{{ \truthset_{ \ifthenelse{\isempty{#1}}{\truth}{{\truth_{#1}}} } }}
\newcommand{\ptruthsubset}[1][]{{ \truthset_{ \ifthenelse{\isempty{#1}}{\truth}{{\truth_{#1}}} }^\possible }}

\newcommand{\xx}[1][]{{ \ifthenelse{\isempty{#1}}{\textcolor{red}{XXX}}{\textcolor{red}{~(XXX {#1} XXX)~}} }}

\newcommand{\lrad}{{L_{rad}}} 
\newcommand{\erad}{{E_{rad}}} 
\newcommand{\liso}{{L_{iso}}}
\newcommand{\eiso}{{E_{iso}}}
\newcommand{\epkz}{{E_{pz}}}
\newcommand{\durz}{{T_{90z}}}
\newcommand{\pbol}{{P_{bol}}}
\newcommand{\sbol}{{S_{bol}}}
\newcommand{\epk}{{E_{p}}}
\newcommand{\dur}{{T_{90}}}

\newcommand{\mz}{{\dot\zeta}}

\defcitealias{hopkins2006normalization}{H06}
\defcitealias{li2008star}{L08}
\defcitealias{butler2010cosmic}{B10}
\defcitealias{lloyd2019comparison}{L19}

\graphicspath{{./}{figures/}}

\title[radio-loud and radio-quiet Gamma-Ray Bursts]{Are there radio-loud and radio-quiet Gamma-Ray Bursts?}

\author[J. Osborne, F. Bagheri and A. Shahmoradi]{
    Joshua A. Osborne $^{1}$\thanks{E-mail: joshua.osborne@uta.edu (JAO)}
    Fatemeh Bagheri$^{1}$\thanks{E-mail: fatemeh.bagheri@uta.edu (FB)}
    Amir Shahmoradi$^{1,2}$\thanks{E-mail: a.shahmoradi@uta.edu (AS) (corresponding author)}
    \\
    $^{1}$Department of Physics, College of Science, The University of Texas, Arlington, TX 76010, USA \\
    $^{2}$Data Science Program, College of Science, The University of Texas, Arlington, TX 76010, USA \\
}

\pubyear{2020}

\begin{document}

    \label{firstpage}
    \pagerange{\pageref{firstpage}--\pageref{lastpage}}
    \maketitle

    \begin{abstract}
        The potential existence of two separate classes of Long-duration Gamma-Ray Bursts (LGRBs) with and without radio afterglow emission, corresponding to radio-bright/loud and radio-dark/quiet populations, has been recently argued and favored in the GRB literature. The radio-quiet LGRBs have been found to have, on average, lower total isotropic gamma-ray emissions ($E_{iso}$) and shorter intrinsic prompt gamma-ray durations (e.g., $T_{90z}$). In addition, a redshift $-T_{90z}$ anti-correlation has been discovered among the radio-loud LGRBs, which is reportedly missing in the radio-quiet class.
        Here we discuss the significance of the differences between the energetics and temporal properties of the two proposed classes of radio-loud and radio-quiet LGRBs. We show that much of the proposed evidence in support of the two distinct radio populations of LGRBs can be explained away in terms of selection effects and sample incompleteness. Our arguments are based on the recent discovery of the relatively-strong highly-significant positive correlation between the total isotropic emission ($E_{iso}$) and the intrinsic prompt duration ($T_{90z}$) that is present in both populations of short-hard and long-soft GRBs, predicted, quantified, and reported for the first time by Shahmoradi (2013) and Shahmoradi \& Nemiroff (2015).
    \end{abstract}


    \section{Introduction}
\label{sec:intro}

    Gamma-ray bursts (GRBs) are extremely energetic explosions, the long-duration class of which has been long hypothesized to be due to the death super-massive stars, releasing energies on the orders of $10^{48}$ to $10^{52}$ ergs \citep[e.g.,][]{frail2001beaming, cenko2010collimation, shahmoradi2015short} and occur at distances far from us on the cosmological scale \citep{metzger1997spectral}. The current afterglow model is that of an expanding fireball \citep{piran1999gamma, meszaros2002theories, woosley2006supernova} where the high energy prompt gamma-ray emission is first observed, followed by afterglow emission at lower-energy frequencies after a few hours or days from the initial gamma-ray prompt emission \citep{frail1997radio, van1997transient, heng2008direct}.
    \newpar

    The standard fireball model used today was proposed after the first observation of both X-ray and optical afterglows in February of $1997$ \citep{kumar2015physics}. Not long after, the first GRB {\it without} an optical afterglow counterpart was found \citep{groot1998search}. This class of bursts without optical afterglows became known as the {\it dark GRBs} \citep{fynbo2001detection}. On October 9, 2000, however, the second High Energy Transient Explorer (HETE-2) \citep{ricker2003high} was launched and in December of 2002 it viewed it's first dark GRB where the optical afterglow was seen, but disappeared and was no longer visible after 2 hours. This begged the question, is there ever truly such a phenomenon as a {\it dark GRB} or is it possible that with better equipment and timing no such phenomena would ever appear?
    \newpar

    More recently, studies have raised the possibility of the existence of a new population of Long-duration GRBs (LGRBs) that are intrinsically dark in the radio-bandwidth afterglow. These events, named `radio-dark', `radio-quiet', or `radio-faint' LGRBs, have been hypothesized to have progenitors that are different from the progenitors of the mainstream `radio-bright' or synonymously-named `radio-loud' LGRBs \citep[e.g.,][]{hancock2013two, lloyd2017lack, lloyd2019comparison}.
    \newpar

\begin{figure}
        \centering
        \includegraphics[width=0.47\textwidth]{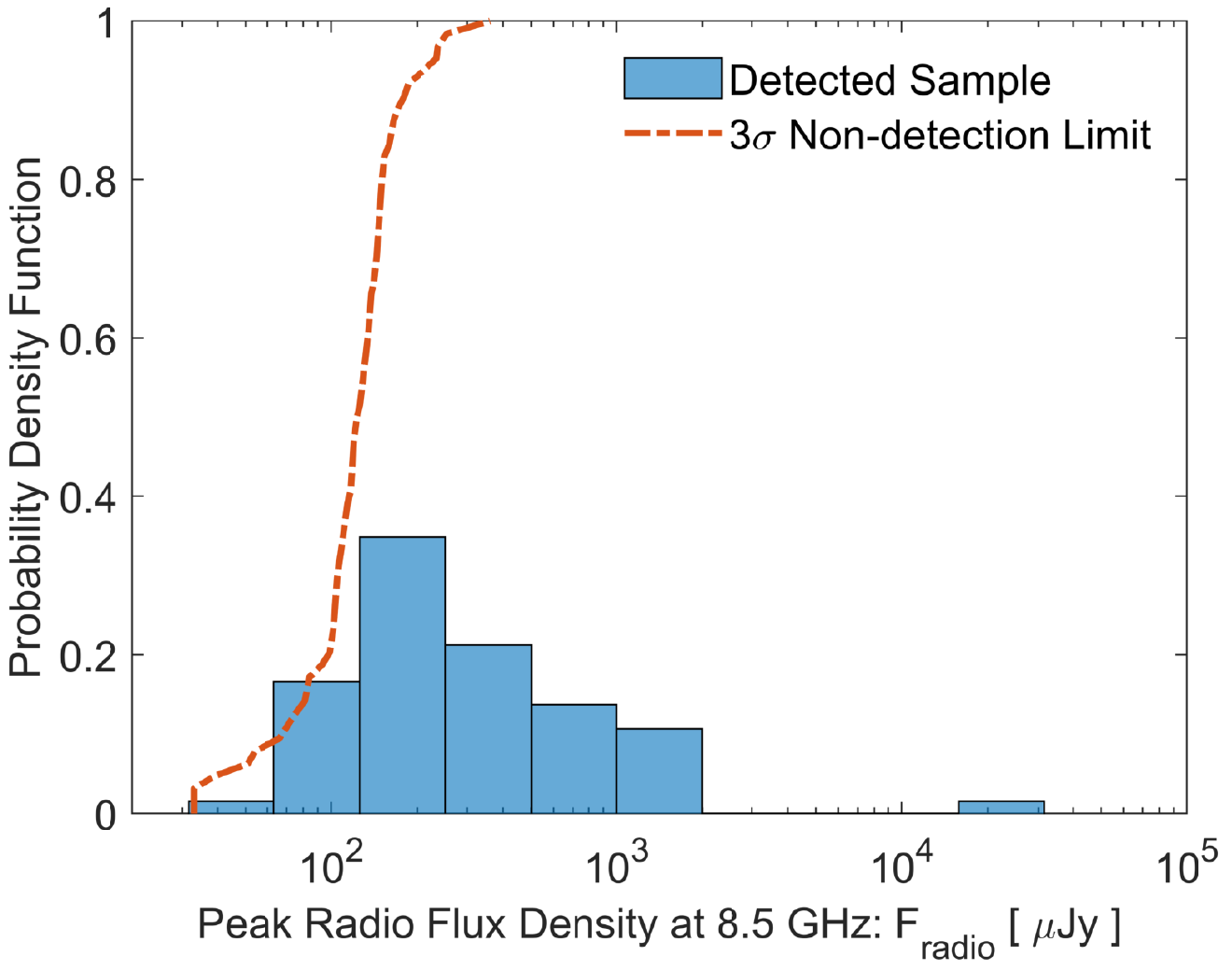}
        \caption{Histogram of the normalized count (probability density function) of 66 radio-loud GRBs taken from \citet{chandra2012radio}. The dashed-dotted red line represents the Cumulative Probability Density Function of the $3\sigma$ upper-limits for the non-detection of radio afterglows of 107 radio-quiet GRBs. \label{fig:ChandraRadioDetectionEfficiency}}
    \end{figure}

    Alternative hypotheses on the lack of detection of radio afterglows in some LGRBs have been also studied \citep[e.g.,][]{frail2005radio, chandra2012radio}. In particular, \citet{chandra2012radio} used a comprehensive sample of 304 afterglows, consisting of 2995 flux density measurements at a wide range of frequencies between 0.6 GHz and 660 GHz, to argue against the potential existence of radio-bright and radio-dark LGRBs. The argument therein is based on the observation that the $3\sigma$ upper-limit for the non-detection of the radio afterglow of radio-dark LGRBs closely traces the faint-tail of the radio-bright LGRB sample, as illustrated in Figure \ref{fig:ChandraRadioDetectionEfficiency}.
    \newpar

    Responding to these observations, \citet{hancock2013two} employed image stacking techniques to increase the overall signal in the combined data from all radio-quiet GRBs. Based on their own specific classification, they conclude that radio-dark LGRBs are on average 2-3 orders of magnitude fainter than radio-bright LGRBs. Such result is not surprising and is in fact, relatively consistent with the observational sample of \citet{chandra2012radio}, illustrated in Figure \ref{fig:ChandraRadioDetectionEfficiency}.
    \newpar

    Nevertheless, based on simulations of the radio afterglows of LGRBs, \citet{hancock2013two} predict that the expected stacked radio flux density of radio-dark LGRBs must be on average 5 times brighter in order to be consistent with the hypothesis of a common continuous unimodal distribution for the radio afterglow properties of all LGRBs together. However, they acknowledge several limitations of their simulations, most importantly, the assumption that the observed sample of radio-afterglows in their studies is representative of the radio afterglow properties of the entire underlying population of detected and undetected LGRBs.
    \newpar

    \begin{figure*}
        \centering
        \makebox[\textwidth]
        {
            \begin{tabular}{ccc}
                \subfloat[]{\includegraphics[width=0.31\textwidth]{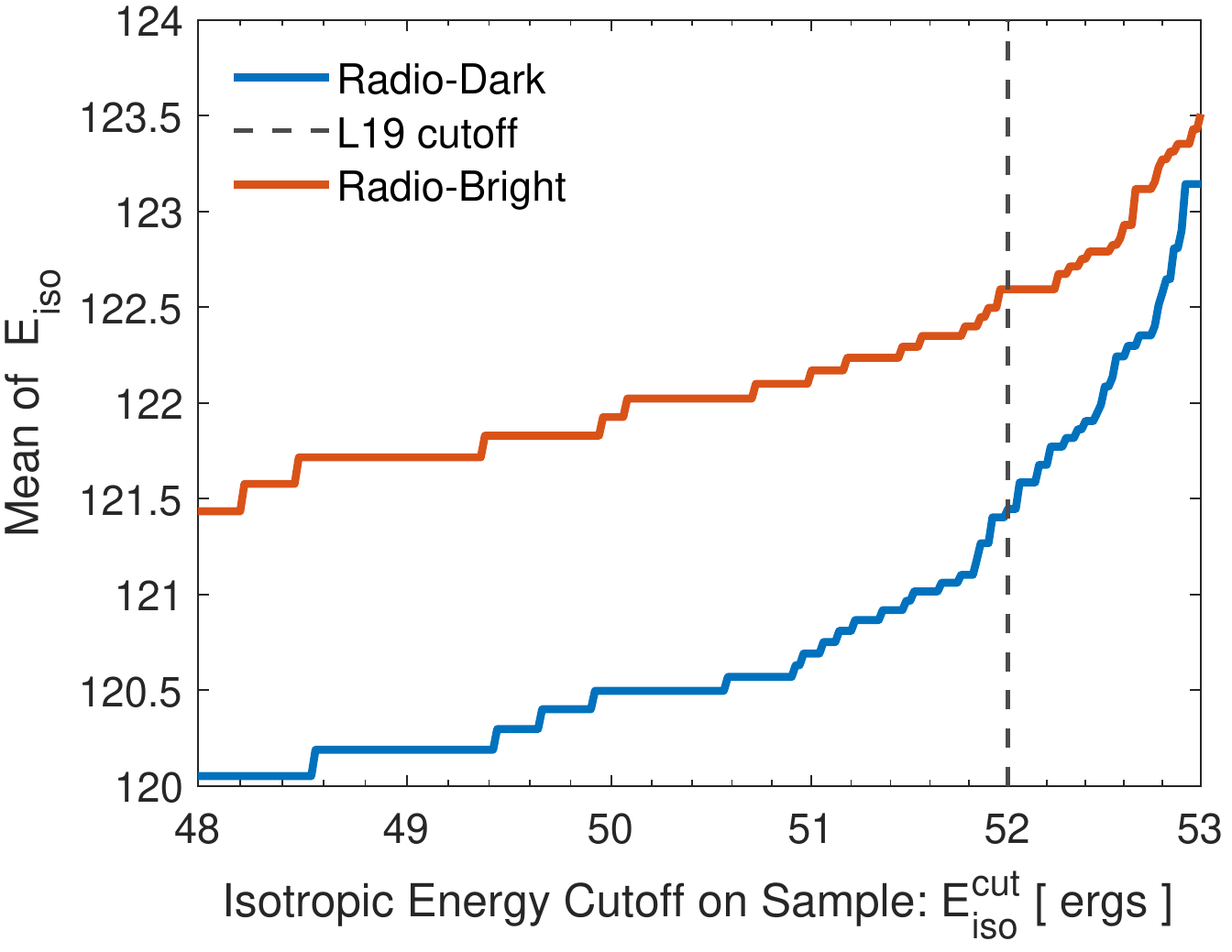} \label{fig:curoffMeanEiso}} &
                \subfloat[]{\includegraphics[width=0.31\textwidth]{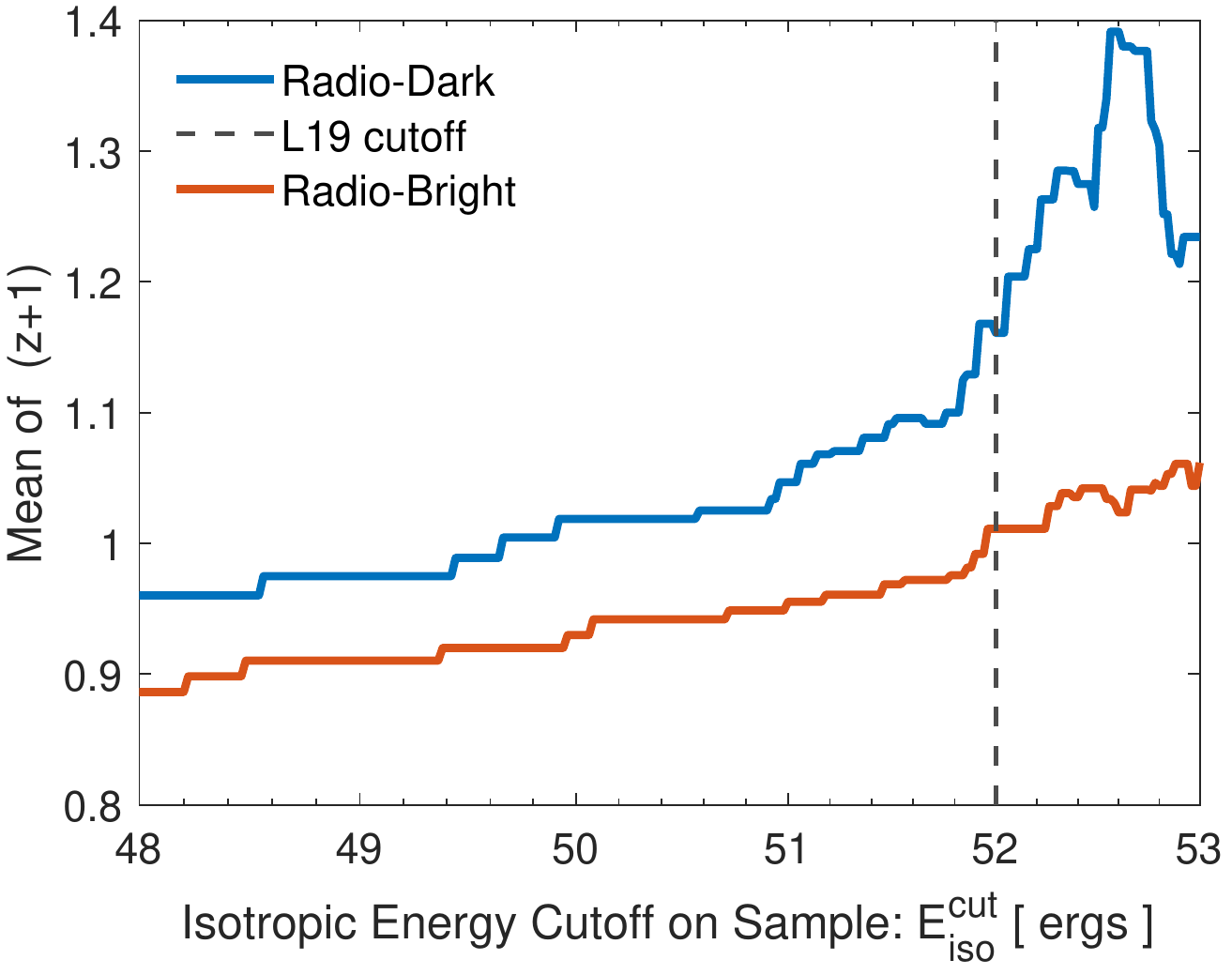} \label{fig:curoffMeanZone}} &
                \subfloat[]{\includegraphics[width=0.31\textwidth]{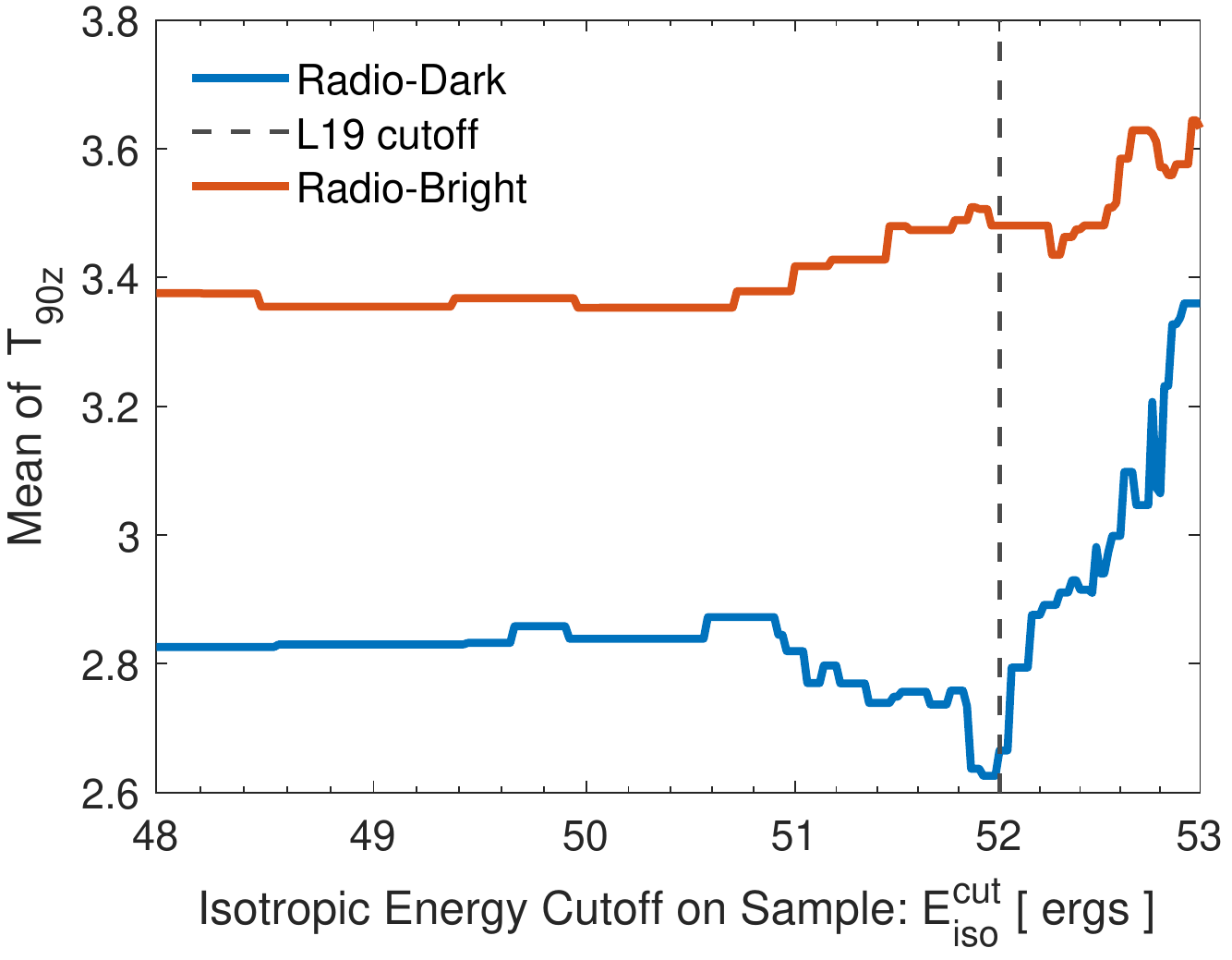} \label{fig:curoffMeanZone}} \\
                \subfloat[]{\includegraphics[width=0.31\textwidth]{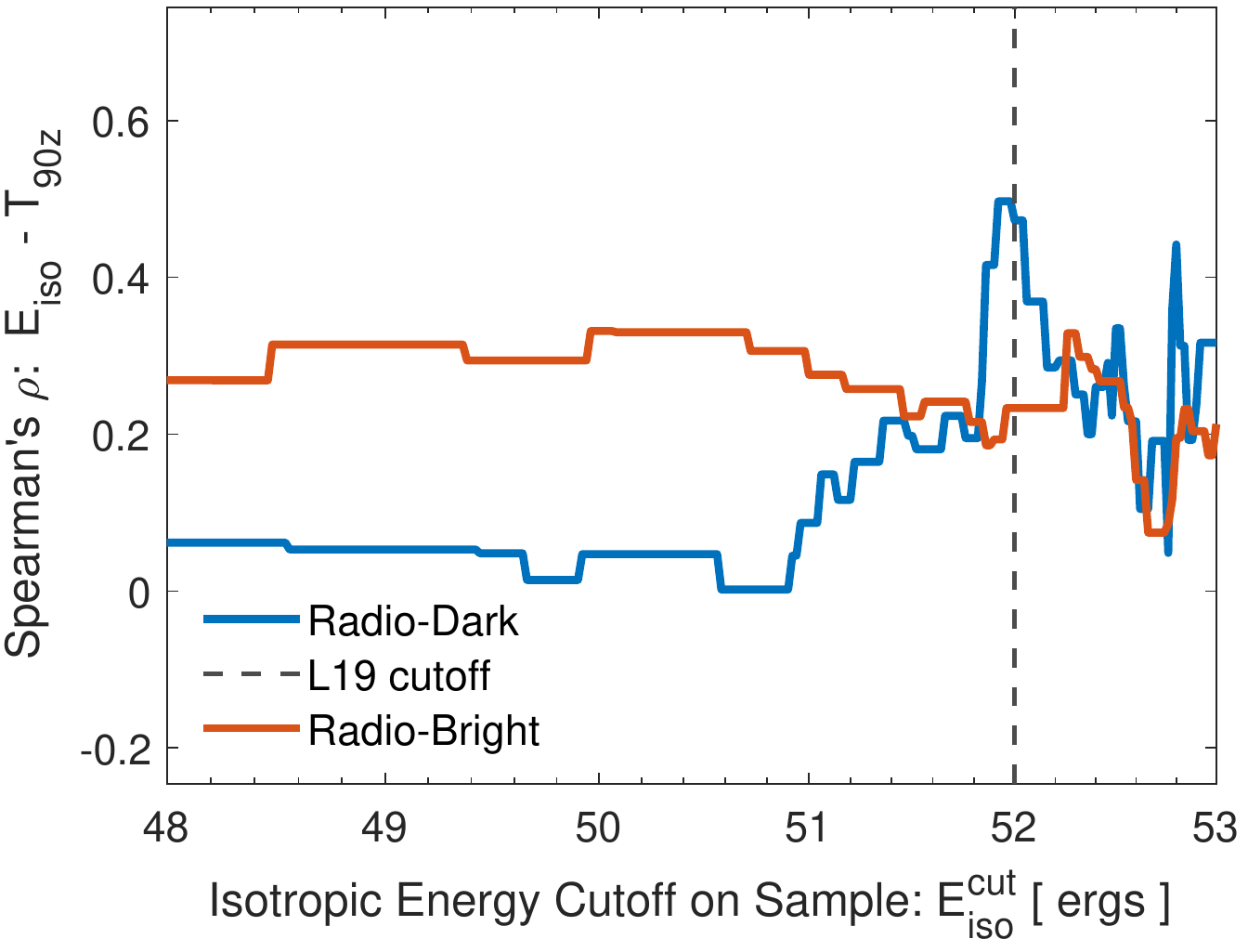} \label{fig:curoffEisoDurz}} &
                \subfloat[]{\includegraphics[width=0.31\textwidth]{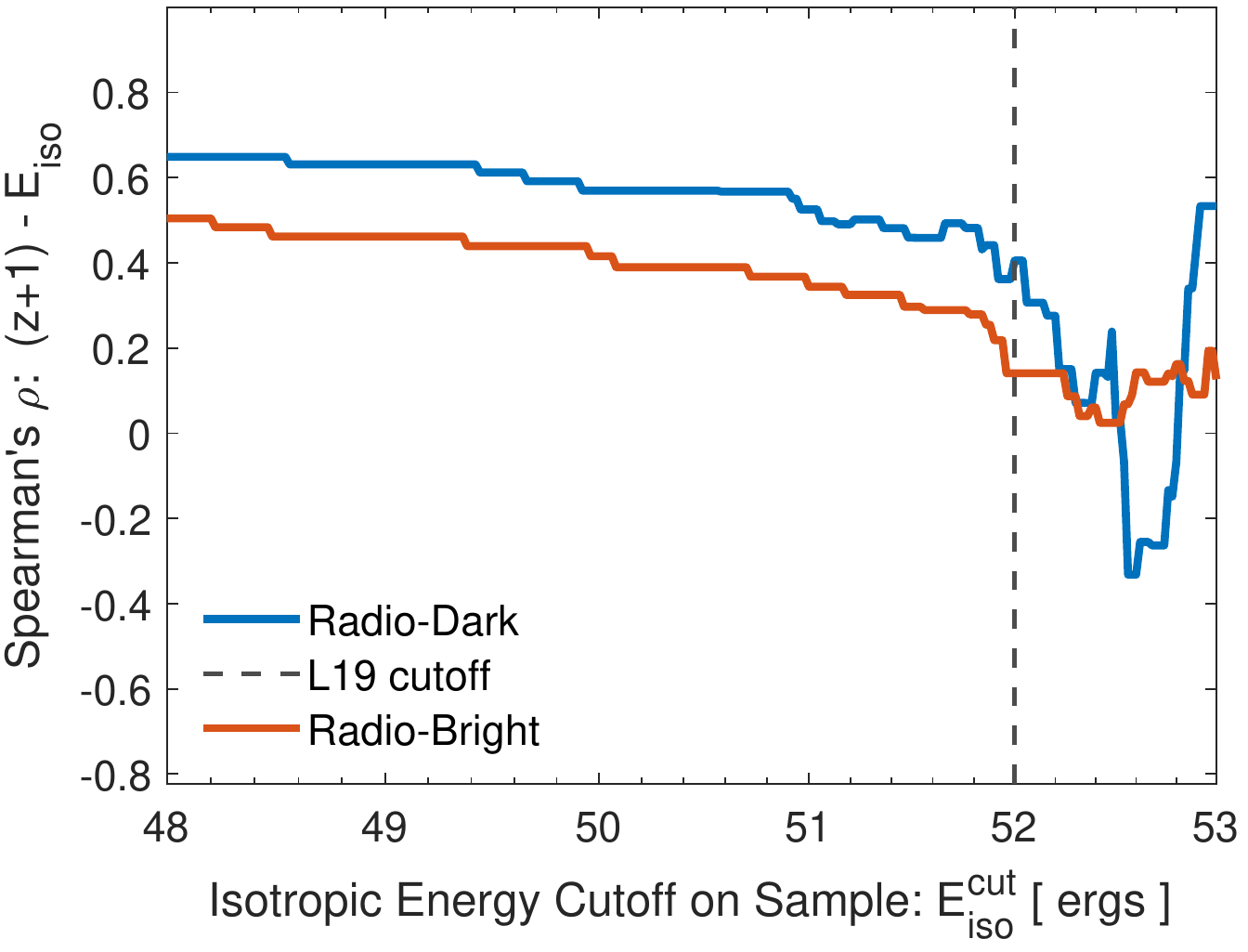} \label{fig:curoffZoneEiso}} &
                \subfloat[]{\includegraphics[width=0.31\textwidth]{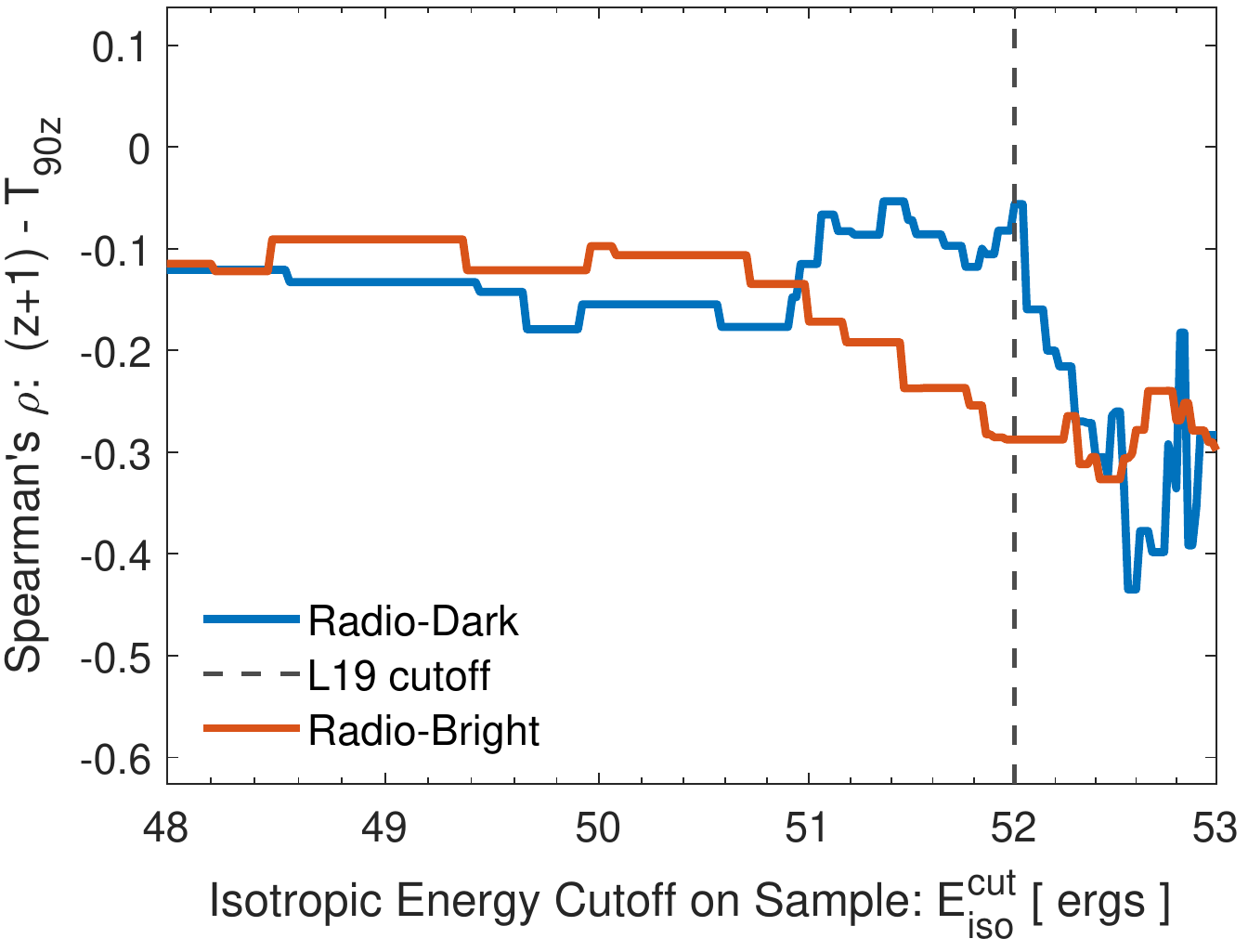} \label{fig:curoffZoneDurz}} \\
            \end{tabular}
        }
        \caption{
        A depiction of the dynamics of different properties of the radio-loud and radio-quiet LGRBs as a function of the sample selection cutoff. The dashed line represent the cutoff used in the study of \citetalias{lloyd2019comparison} to generate a sample of bright LGRBs. As evidenced, some of the differing attributes of the two radio classes appear to be highly sensitive to the arbitrarily chosen sample selection cutoff value used in \citetalias{lloyd2019comparison}, most importantly, the correlations between redshift ($z$), isotropic emission ($\eiso$), and intrinsic duration ($\durz$).
        \label{fig:VaryingCutoffs}
        }
    \end{figure*}

    Most recently, \citet{lloyd2019comparison} (hereafter \citetalias{lloyd2019comparison}) build upon the existing sample of the GRB radio afterglows of \citet{chandra2012radio} to further confirm and strengthen the findings of their previously published study \citep[][]{lloyd2017lack} on the fundamental spectral and temporal differences and similarities of the radio-quiet and radio-loud GRBs in multiple observational bandwidths including optical, x-ray, gamma-ray and GeV emission. Specifically, they find that,
    \begin{enumerate}
        \item
            The total isotropic gamma-ray emission ($\eiso$) does not correlate with the radio luminosity in their observational samples of 78 radio-loud and 41 radio-quiet LGRBs.
        \item
            Radio-quiet LGRBs have significantly lower total isotropic gamma-ray emission ($\eiso$), on average, 5 times lower than the radio-loud LGRBs.
        \item
            Radio-quiet LGRBs have significantly shorter intrinsic prompt duration as measured by the quantity $\durz$ \citep[see for example,][]{shahmoradi2013multivariate, shahmoradi2015short}, than the radio-loud LGRBs.
        \item
            Radio-quiet LGRBs exhibit a weak positive $\eiso-\durz$ correlation whereas this correlation is missing in the radio-loud LGRBs.
        \item
            Radio-quiet LGRBs exhibit a weak negative correlation between $\durz$ and redshift ($z$) whereas the radio-loud LGRBs exhibit a much stronger such correlation.
        \item
            The very high energy ($0.1-100$ GeV) extended emission is only present in the radio-loud sample.
        \item
            Also, there is no significant difference in the the redshift distribution, the presence of X-ray/optical plateaus, or the average jet opening angles between the two radio-quiet and radio-loud GRB samples.
    \end{enumerate}

    Such observational evidence in favor of the two classes of radio-loud and radio-quiet LGRBs can have profound implications for the progenitors of LGRBs as highlighted and discussed by \citetalias{lloyd2019comparison}. However, here we argue and show that much of the above evidence in favor of the two fundamentally-distinct radio-quiet and radio-loud GRBs can be potentially explained in terms of the existing correlation between $\eiso$ and $\durz$ of both long and short duration classes of GRBs. To the extent of our knowledge, this positive $\eiso-\durz$ correlation in both classes of LGRBs and short-duration GRBs (SGRBs) was originally discovered, quantified, and reported for the first time by \citet{shahmoradi2013multivariate} and \citet{shahmoradi2015short} via a careful analysis of the largest catalog of GRBs available at the time. Hints to the existence of such a correlation has been also independently provided by \citet{butler2010cosmic}. The relationship has been also rediscovered recently in an independent study of a sample of Swift-detected GRBs \citep[][]{tu2018correlation}.
    \newpar

    In the following sections, we consider the arguments in favor of the existence of two separate populations of radio-loud and radio-quiet GRBs and show that much of the evidence provided can be likely explained in terms of strong selection effects and sample incompleteness of the radio afterglow data and the existing correlations among the spectral and temporal properties of LGRBs as discovered and quantified by \citet{shahmoradi2013multivariate, shahmoradi2015short}.

    \section{An alternative interpretation for the existence of radio loud and quiet LGRBs}
\label{sec:methods}

    \begin{figure*}
        \centering
        \makebox[\textwidth]
        {
            \begin{tabular}{ccc}
                \subfloat[]{\includegraphics[width=0.31\textwidth]{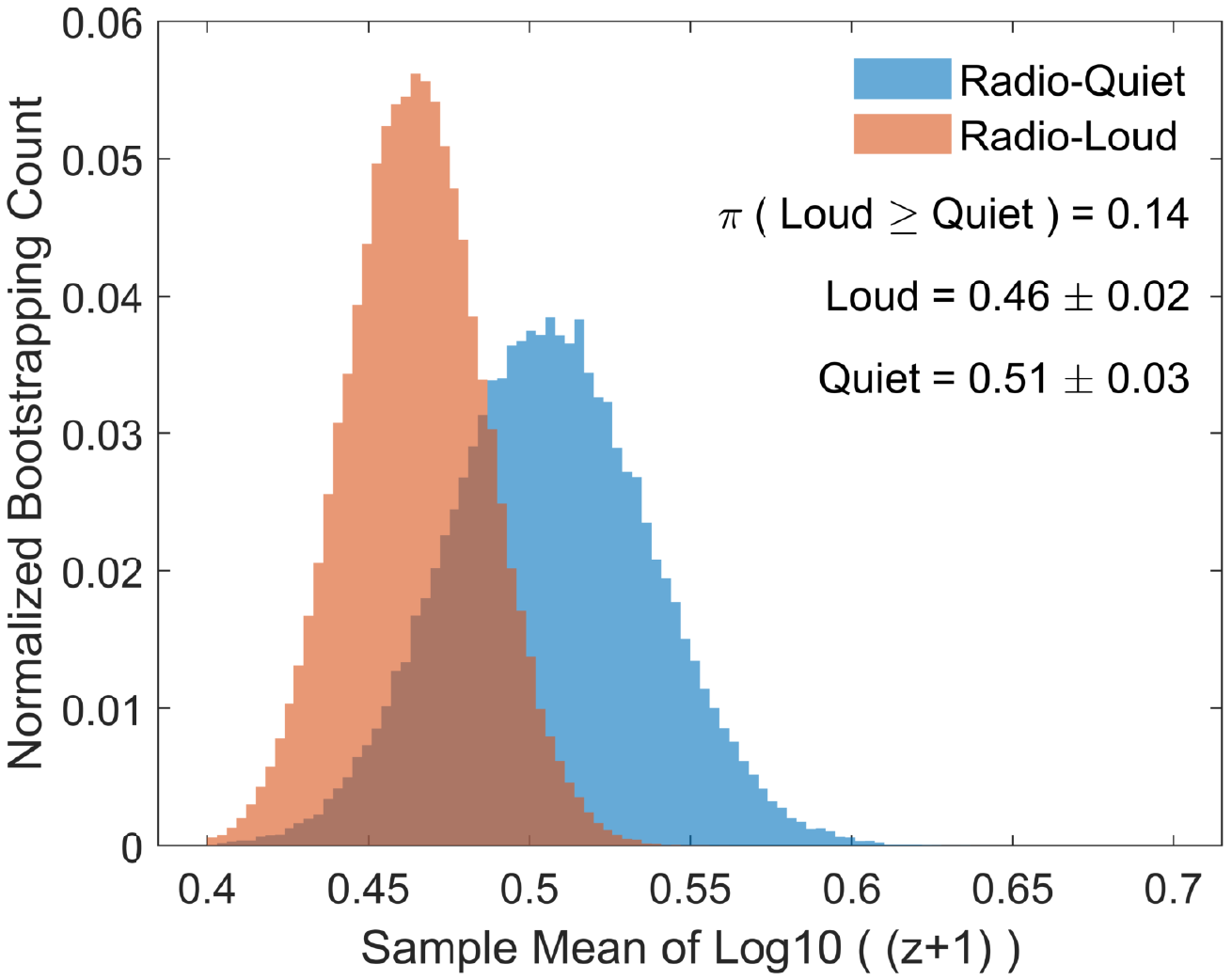} \label{fig:bootL19a}} &
                \subfloat[]{\includegraphics[width=0.31\textwidth]{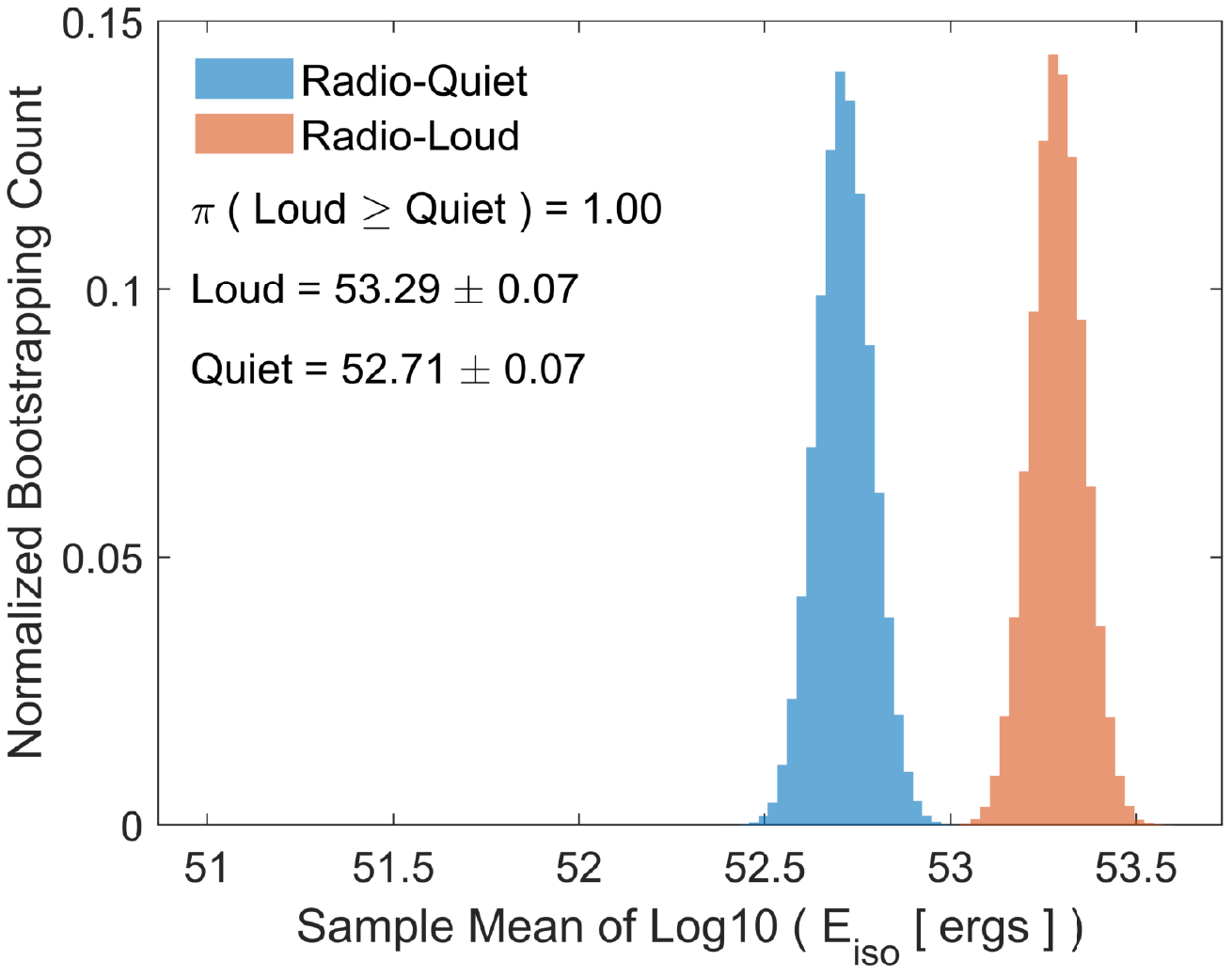} \label{fig:bootL19b}} &
                \subfloat[]{\includegraphics[width=0.31\textwidth]{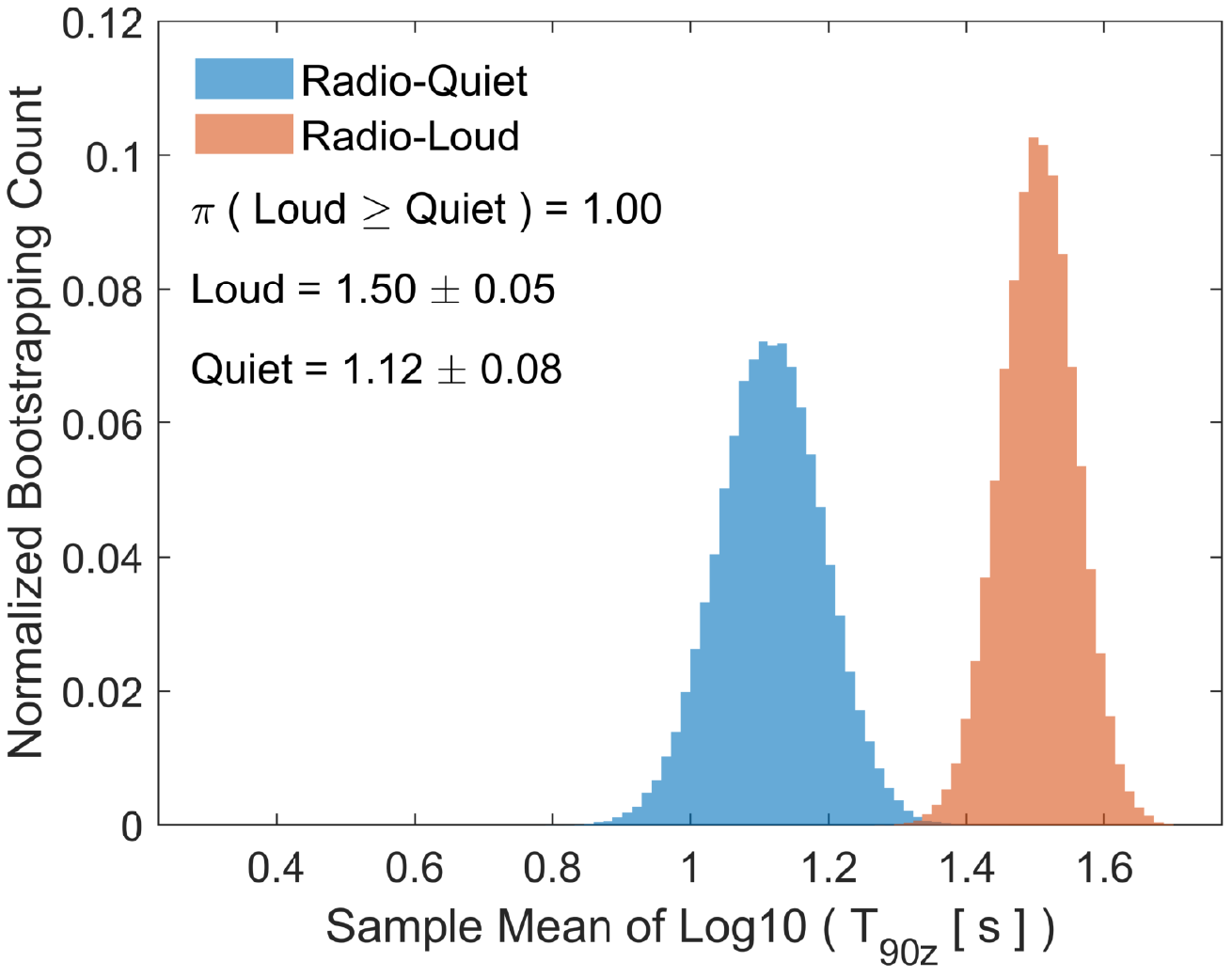} \label{fig:bootL19c}} \\
                \subfloat[]{\includegraphics[width=0.31\textwidth]{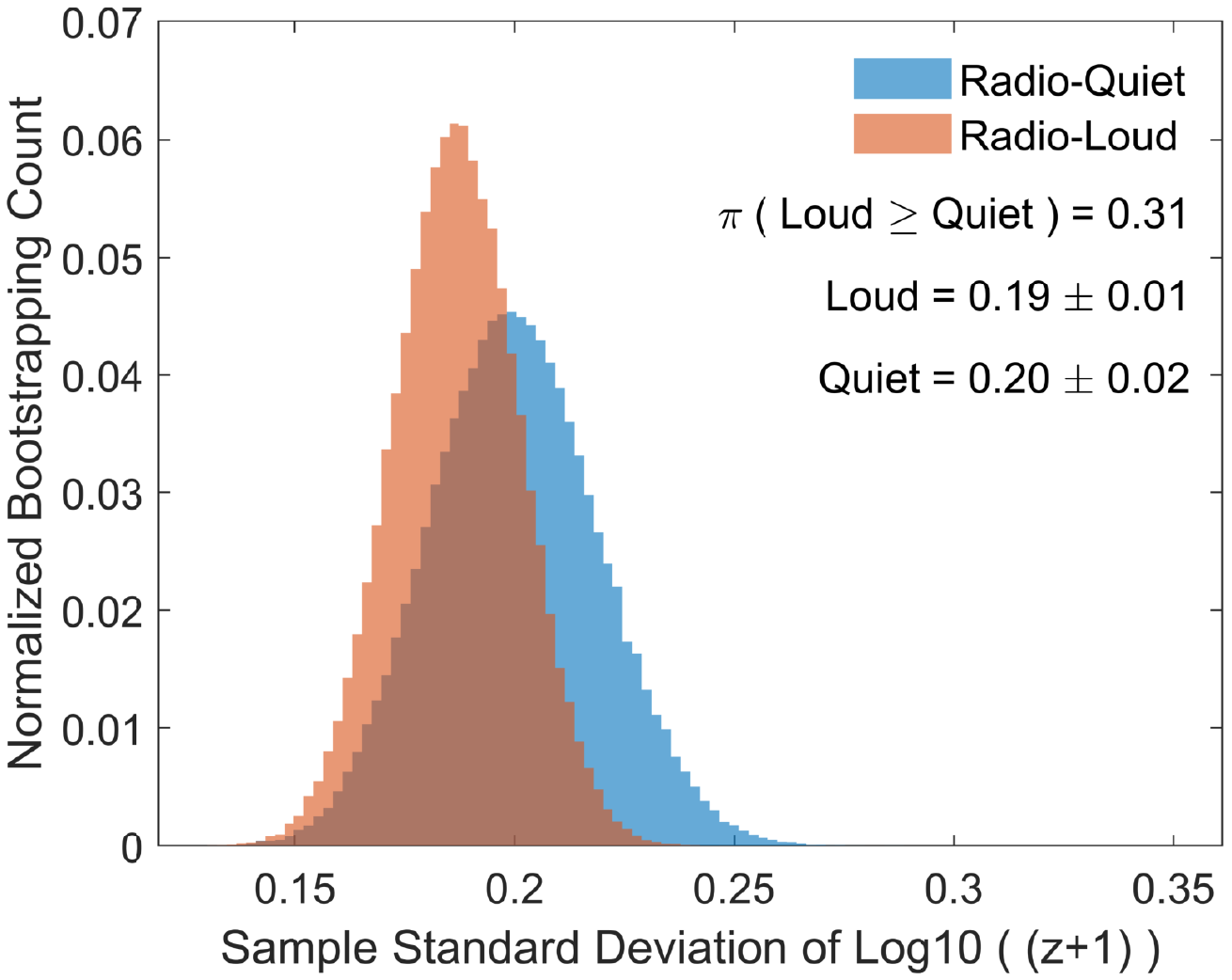} \label{fig:bootL19d}} &
                \subfloat[]{\includegraphics[width=0.31\textwidth]{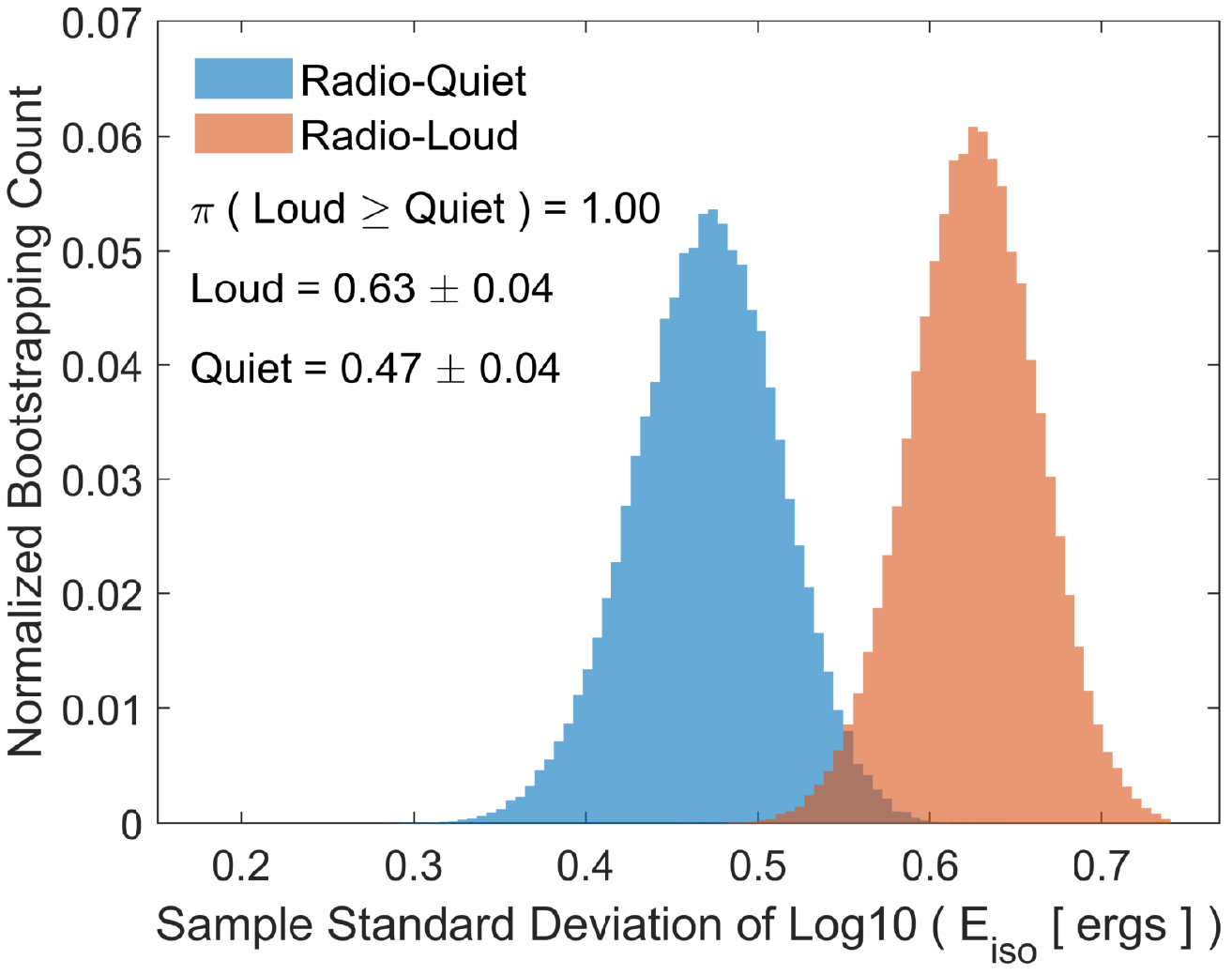} \label{fig:bootL19e}} &
                \subfloat[]{\includegraphics[width=0.31\textwidth]{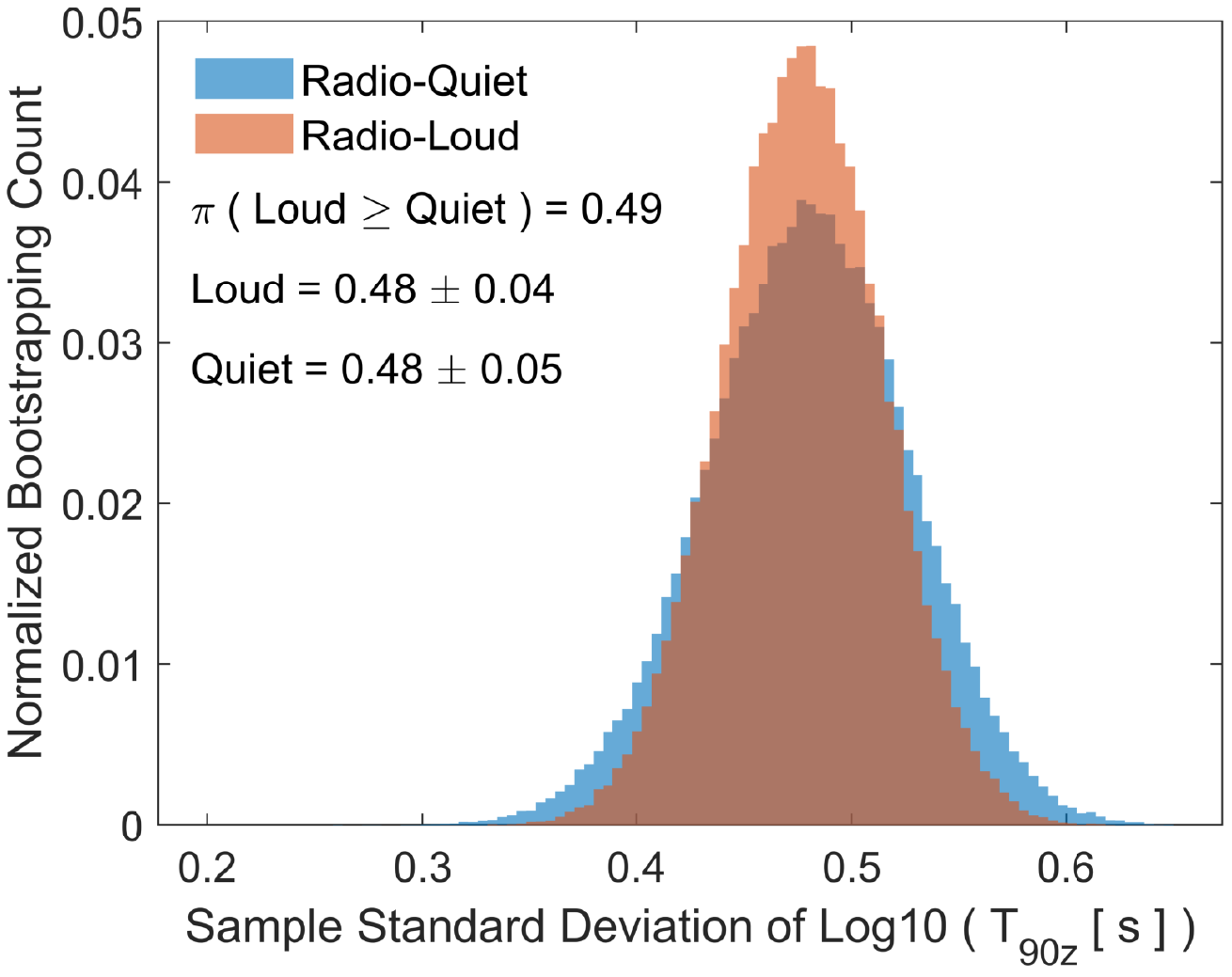} \label{fig:bootL19f}} \\
                \subfloat[]{\includegraphics[width=0.31\textwidth]{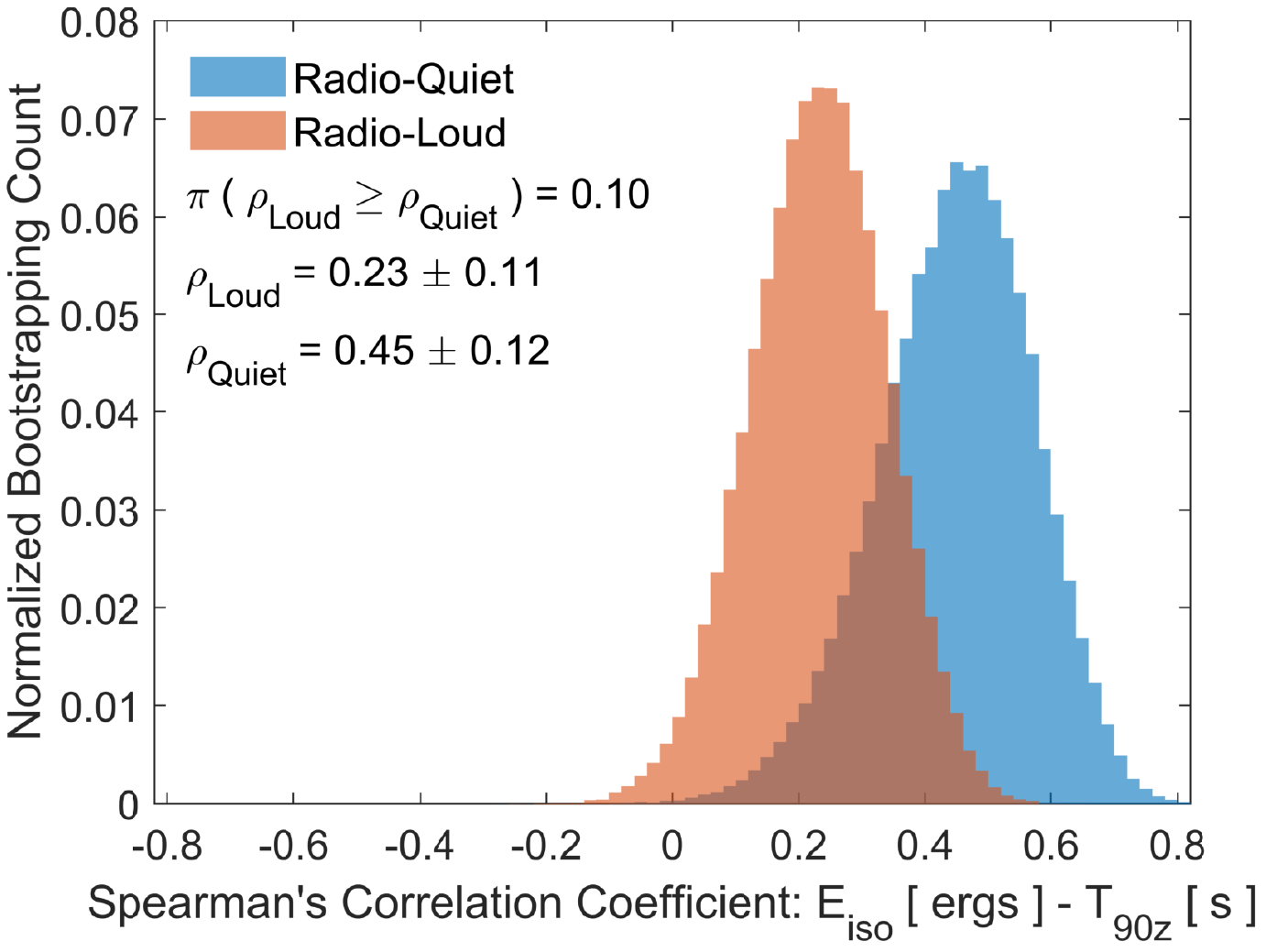} \label{fig:bootL19g}} &
                \subfloat[]{\includegraphics[width=0.31\textwidth]{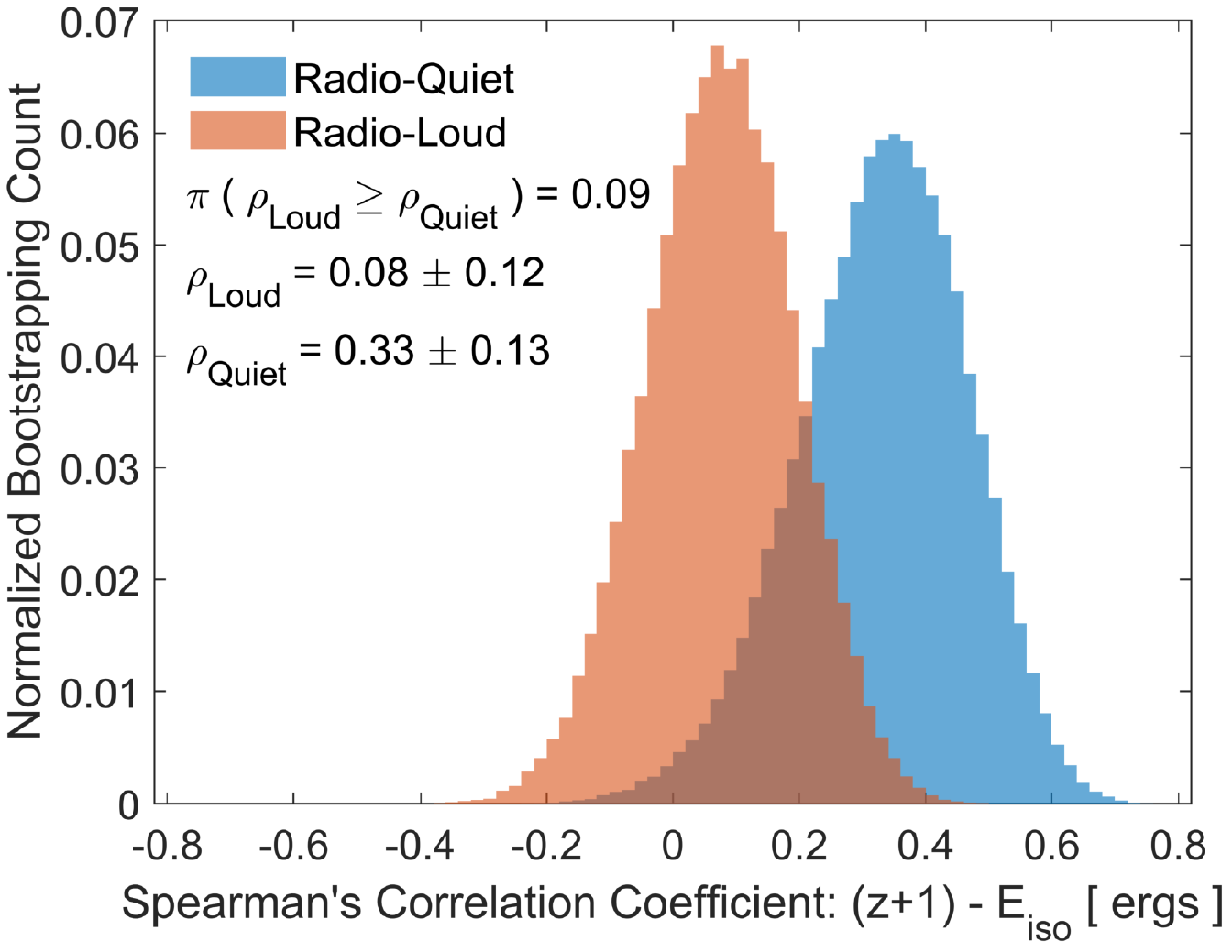} \label{fig:bootL19h}} &
                \subfloat[]{\includegraphics[width=0.31\textwidth]{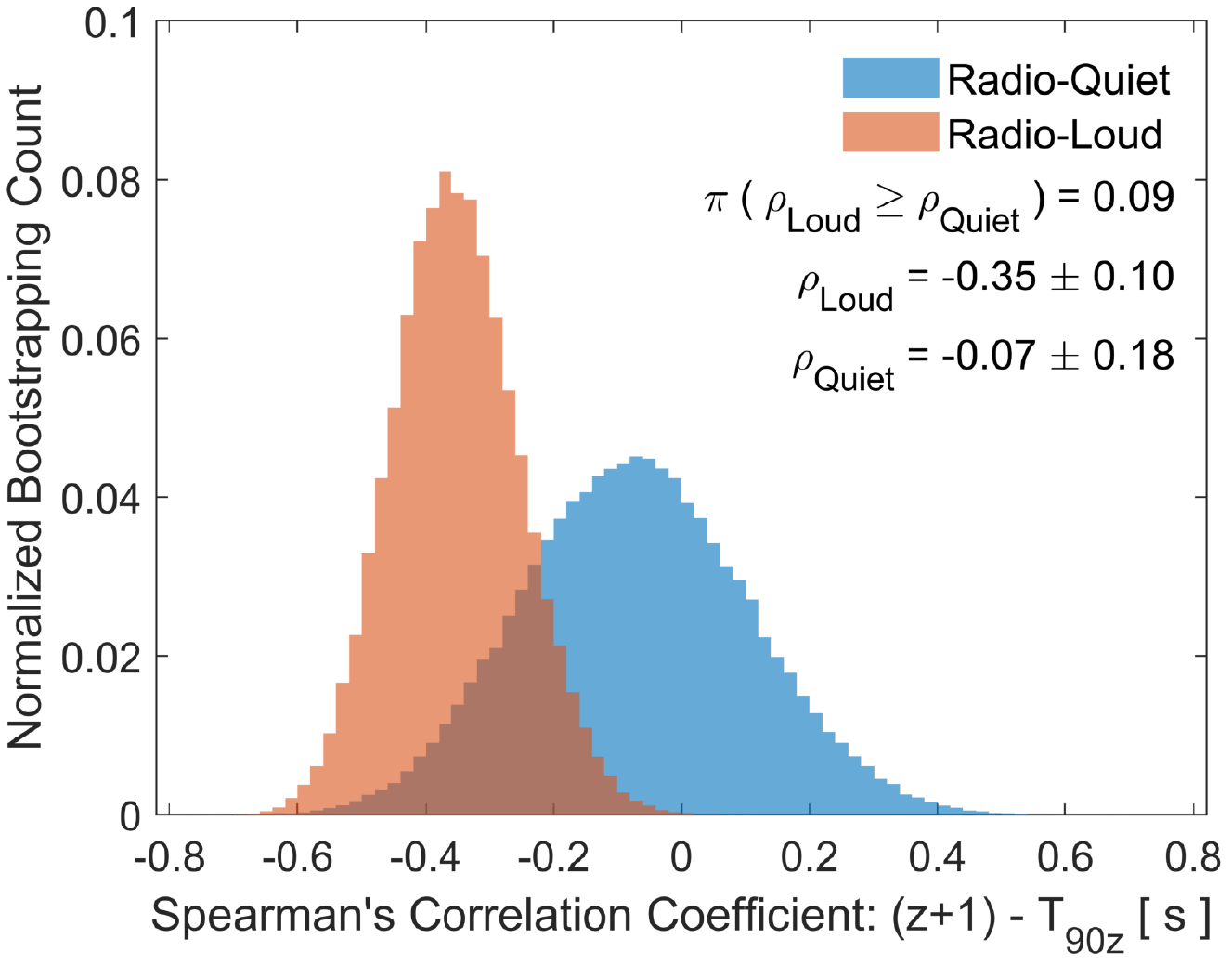} \label{fig:bootL19i}} \\
            \end{tabular}
        }
        \caption{
        The bootstrapping simulation results for the some of the population properties of the two classes of radio-loud and radio-quiet LGRBs. Plots (a), (b), (c) display respectively, the bootstrapping distributions of the averages of the distributions of $z+1$, $\eiso$, $\durz$ distributions of the two radio classes. In the same order, plots (d), (e), (f) display the corresponding bootstrapping distributions of the standard deviations of the three LGRB property distributions for the two classes. Plots (g), (h), (i) display the bootstrapping distributions of the Spearman's correlation strengths between the three LGRB properties. The inset values represent the mean and the $1\sigma$ dispersions in the bootstrap distributions in each plot, with $\pi(\cdot)$ denoting the probability and $\rho$ denoting the Spearman's correlation strength.
        \label{fig:bootL19}
        }
    \end{figure*}

    To better quantify the differences between the two hypothesized radio classes, we first apply the statistical bootstrapping \citep[e.g.,][]{efron1994introduction} technique to the radio-loud and radio-quiet samples of \citetalias{lloyd2019comparison} to form confidence bounds on the mean and standard deviations of the two populations, as well as the correlation strengths between the three quantities: redshift ($z$), the total isotropic gamma-ray emission ($\eiso$), and the intrinsic duration ($\durz$). In brief, bootstrapping is a statistical resampling method (with replacements) that, in the absence of multiple independent sets of observational data, can provide confidence bounds on the various statistical properties of a dataset.
    \newpar

     In the following sections, we discuss and quantify the significance of each of the differences and similarities of the two radio-loud and radio-quiet LGRBs as quantified by the bootstrap confidence bounds and provide further evidence from Monte Carlo simulations that shed light on the classification of the radio emissions of LGRBs.

    \subsection{The redshift distributions of radio-loud and radio-quiet LGRBs}
    \label{sec:methods:energetics}

        Figure \ref{fig:bootL19} displays the bootstrapping results for some of the statistical properties of the two radio classes of \citetalias{lloyd2019comparison}. The first and foremost quantity in the studies of cosmological objects is redshift ($z$). Similar to \citetalias{lloyd2019comparison}, we find the redshift distributions of the two classes of radio-loud and radio-quiet LGRBs are consistent with hypothesis of belonging to the same parent population distribution. A two-sample Kolmogorov-Smirnov test on the two redshift distributions reveals no significant difference between the two redshift distributions, that is, we cannot reject the null hypothesis that the two redshift samples originate from the same distribution with a KS-test probability of $\rho\sim0.24$.
        \newpar

        We note, however, that this hypothetical parent distribution for the redshifts of the two classes is likely severely affected by selection effects due to the gamma-ray detection and the redshift measurement thresholds. We find the average redshifts of the radio-quiet and radio-loud samples to be $\overline{z}_{loud}\sim2.21<\overline{z}_{quiet}\sim2.6$. This is contrary to the reported value for $\overline{z}_{loud}$ in Table (1) of \citetalias{lloyd2019comparison}. Nevertheless, our bootstrapping simulations indicate that the differences in the redshift averages are insignificant as illustrated in Figure \ref{fig:bootL19a}. There is a $0.14$ probability that the average redshift of the radio-loud sample could be actually larger than the average redshift of the radio-quiet sample. The dispersion in the redshift distributions of the two radio classes also appear to be consistent with each other as depicted by the bootstrapping results in Figure \ref{fig:bootL19d}.

    \subsection{The energetics of radio-loud and radio-quiet LGRBs}
    \label{sec:methods:energetics}

        \begin{figure*}
            \centering
            \makebox[\textwidth]{
            \subfloat[$\eiso$ vs. Peak Radio Luminosity]{%
                \includegraphics[width=0.47\textwidth]{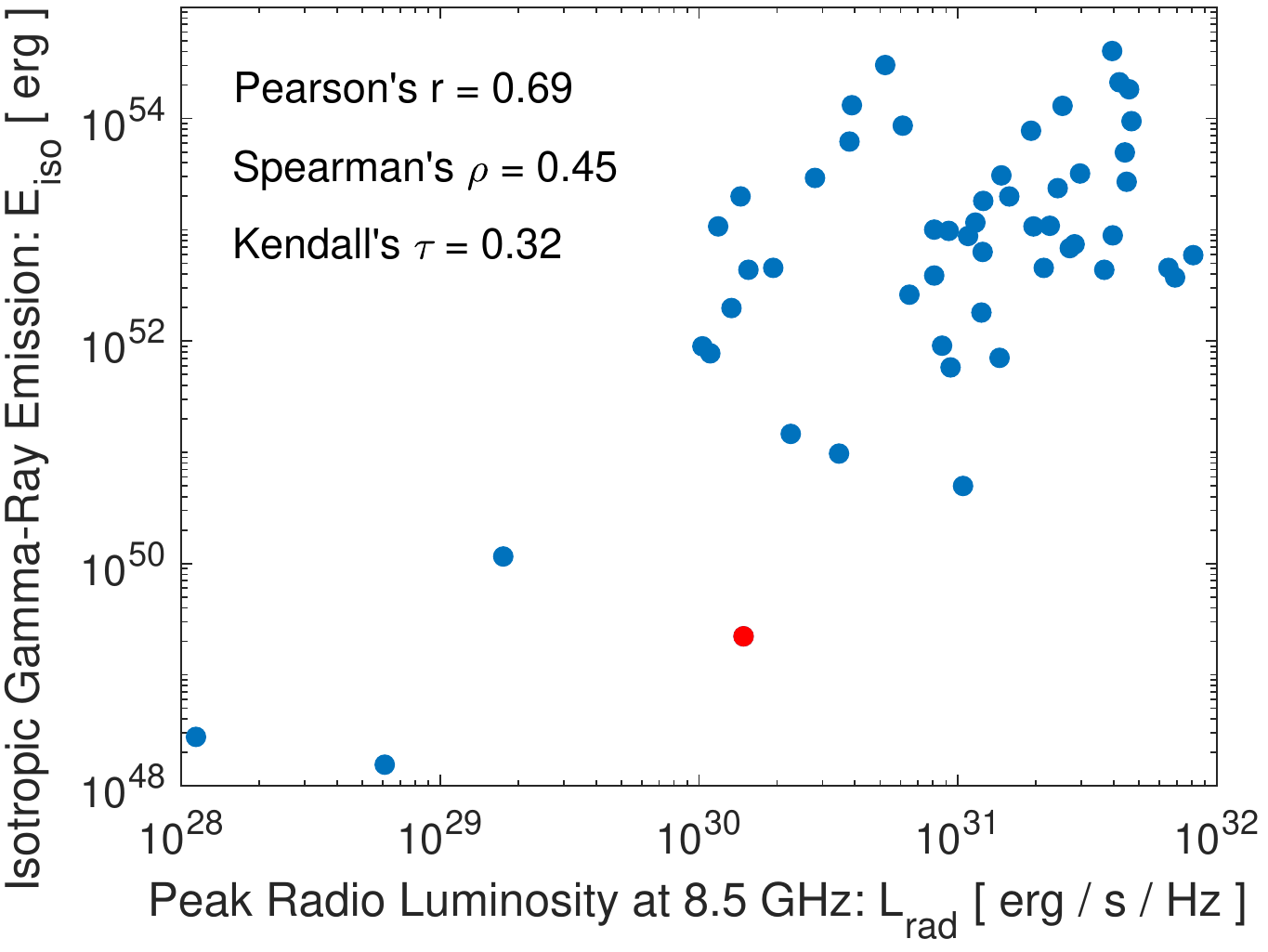}
                \label{fig:ChandraLradEiso}}
            \quad
            \subfloat[A Depiction of detector thresholds and classification limit]{%
                \includegraphics[width=0.47\textwidth]{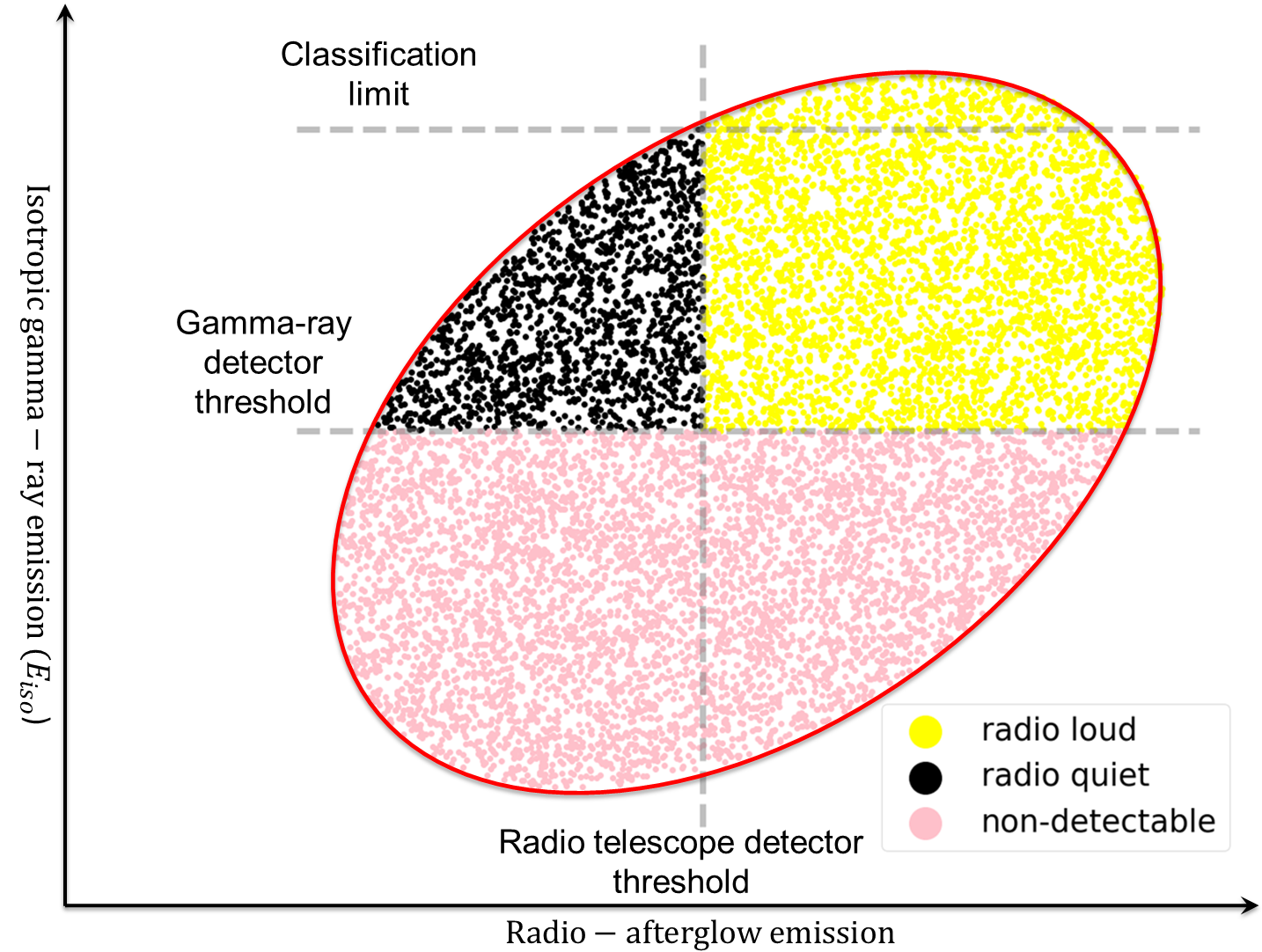}
                \label{fig:RadioCutoffs}}
                }
            \caption
            {
                {\bf (a)}: An illustration of the observational sample of \citet{chandra2012radio} showing the relationship between $\eiso$ and the peak radio luminosity. The only X-Ray Flash (XRF) event in the sample of \citet{chandra2012radio} is shown by the red point. Exclusion of the XRF from the plots results in a minor decrease in the Spearman's correlation strength from $\rho=0.45\pm0.13$ to $\rho=0.43\pm0.13$.
                {\bf (b)}: A schematic illustration of the combined effects of the gamma-ray and radio detection thresholds as well as the potential existence of an underlying correlation between the radio and gamma-ray emissions of LGRBs on the observed LGRBs sample and how these factors can lead to the appearance of two separate classes of radio-loud and radio-quiet LGRBs. The red oval represents the positive underlying correlation between $\eiso$ and the radio emission. However, the gamma-ray detector threshold (and the complex redshift selection effects) impose a severe cut on the y-axis, hiding much of the cosmic population of LGRBs (represented by the pink color) from our view. Simultaneously, the radio detector threshold effects, place another severe cut on the observed sample of LGRBs to create an impression of two separate classes of radio-loud and radio-quiet LGRBs. Because of the positive correlation of $\eiso$ with the radio emission, any LGRB above a certain $\eiso$ threshold (i.e., the {\it radio classification limit}) will be automatically {\it always} classified as a radio-loud event. This leads to an apparent increase in the average $\eiso$ of the radio-loud sample (represented by the yellow-colored points) relative to the radio-quiet sampled (represented by the black-colored points), similar to what has been reported in the literature as the differing characteristics of radio-quiet and radio-loud LGRBs.
                \label{fig:chandraSchematic}
            }
        \end{figure*}

        Similar to \citetalias{lloyd2019comparison}, the results of our bootstrapping simulations indicate a strong evidence, at almost $100\%$ confidence level, that the average and the standard deviation of the $\eiso$ distribution of the radio-loud LGRBs are larger than the average and the standard deviation of the $\eiso$ distribution of the radio-quiet sample (Figures \ref{fig:bootL19b} \& \ref{fig:bootL19e}). However, unlike \citetalias{lloyd2019comparison}, here we hypothesize a different origin for the observed differences in the $\eiso$ distributions of the radio-loud and radio-quiet LGRBs.
        \newpar

        Our {\it fundamental hypothesis} in this work is that {\it there is potentially a significant (but not necessarily strong) positive correlation between the radio-afterglow and the prompt gamma-ray energetics of LGRBs}. Such a hypothesis may not be too far from reality as there has been already evidence for the potential existence of a positive correlation between the total isotropic gamma-ray emission ($\eiso$) of GRBs with their peak radio luminosity at 8.5 GHz ($\lrad$). In Figure \ref{fig:ChandraLradEiso}, we have regenerated the plot of Figure 20 in \citet{chandra2012radio}. We find a positive Spearman's correlation strength of $\rho\sim0.45$ for the $\eiso-\lrad$ relationship.
        \newpar

        The scatter in the underlying intrinsic $\eiso-\lrad$ correlation of the LGRBs population is likely different from what is seen in Figure \ref{fig:ChandraLradEiso} since the distributions of both quantities $\eiso$ and $\lrad$ are severely affected by the corresponding gamma-ray and radio detector thresholds. We defer a quantification of this relationship to a future work and suffice, in this work, to only note the high plausibility of the validity such hypothesis given the existing evidence. Hints to the existence of correlations between the prompt gamma-ray and afterglow emissions in wavelengths other than radio have been also provided by other independent studies \citep[e.g.,][]{margutti2013prompt, dainotti2017gamma}.
        \newpar

        The existence of such correlation between the radio and gamma-ray energy releases would readily explain the appearance of the two classes of radio-quiet and radio-loud LGRBs: {\bf The more energetic LGRBs in gamma-ray emission tend to be more luminous in radio afterglows, and therefore, tend to be classified as radio-loud LGRBs more frequently}. Consequently, radio-loud LGRBs appear to be much more energetic as measured by $\eiso$ relative to radio-quiet LGRBs. This phenomenon is well illustrated in the schematic plot of Figure \ref{fig:RadioCutoffs} where the effects of the radio and gamma-ray detector thresholds create apparently two distinct classes of radio loud and quiet LGRBs with significantly different characteristics, similar in behavior to the findings of \citetalias{lloyd2019comparison}.
        \newpar

        As illustrated in Figure \ref{fig:RadioCutoffs}, when LGRBs reach a certain gamma-ray emission as measured by $\eiso$, their radio-emission also surpasses the radio-emission detection threshold. Therefore, LGRBs beyond a certain $\eiso$ threshold are automatically classified as radio-loud LGRBs. This results in the apparent segregation of the LGRB population into two distinct groups whose $\eiso$ distributions are different despite having the same redshift distributions. Indeed, \citetalias{lloyd2019comparison} report such excess in the average energetics of their radio-loud LGRB sample relative to radio-quiet LGRBs (e.g., see the middle plots of Figures 1 \& 2 of \citetalias{lloyd2019comparison} and Figure \ref{fig:bootL19b} in this work). Meanwhile, the presence of two strong and independent radio and gamma-ray detection thresholds on the bivariate $\eiso-\durz$ distribution severely undercuts any traces of a significant $\eiso-\erad$ correlation, where $\erad$ denotes the total radio emission.
        \newpar

        The above alternative hypothesis for the origins of the two radio classes automatically provides also a natural explanation for the significantly smaller dispersion in the $\eiso$ distribution of radio-quiet LGRBs relative to the radio loud sample, which is evident the middle plots of Figures 1 \& 2 of \citetalias{lloyd2019comparison} and Figure \ref{fig:bootL19e} in this work. The $\eiso$ distribution of the radio-quiet sample is strongly affected by the gamma-ray detection threshold and redshift-measurement selection effects in its lower tail and, by the radio-loud classification limit at its upper tail. However, in the case of radio-loud LGRBs, such radio-classification upper limit on the $\eiso$ distribution does not exist and the lower-limit on this distribution is also vaguely defined by a classification limit whose sharpness depends on the strength of the $\eiso-\erad$ correlation. These effects are well-illustrated in Figure \ref{fig:RadioCutoffs}.

    \subsection{The duration distribution of radio-loud and radio-quiet LGRBs}
    \label{sec:methods:durz}

        As soon as the origins of the energetics differences between the radio-loud and radio-quiet LGRBs are understood and accepted as explained in the previous section, the apparent differences between the duration distributions of radio-loud and radio-quiet LGRBs can be also readily explained.
        We do so by utilizing the recent discovery of the potential existence of a strong positive correlation between the gamma-ray energetics of LGRBs and their intrinsic durations, $\eiso-\durz$, quantified for the first time in a series of works by \citet{shahmoradi2013multivariate, shahmoradi2013gamma, shahmoradi2015short, shahmoradi2019catalog, 2019arXiv190306989S, osborne2020multilevel, 2020arXiv200601157O}.
        \newpar

        Plot (b) of Figure \ref{fig:monteCarloUniverse} displays a reproduction of the LGRB world model of \citet{shahmoradi2019catalog, osborne2020multilevel} who find an underlying Pearson's correlation coefficient of $\rho\sim0.5-0.6$ between the intrinsic cosmic distributions of $\eiso$ and $\durz$ in log-log space. Interestingly, \citet{shahmoradi2015short} discover a correlation of similar strength and significance to that of LGRBs in the population of SGRBs.
        \newpar

        Since the radio-loud sample of LGRBs has, on average, higher $\eiso$ than the radio-quiet LGRBs (by about 0.58 dex as illustrated in Figure \ref{fig:bootL19b}, the strong positive $\eiso-\durz$ correlation depicted in plot (b) of Figure \ref{fig:monteCarloUniverse} also necessitates, on average, higher $\durz$ values for radio-loud LGRBs relative to the radio-quiet sample. Such difference has been indeed reported by \citetalias{lloyd2019comparison} (e.g., see the middle plots of Figures 1 \& 2 of \citetalias{lloyd2019comparison} and Figure \ref{fig:bootL19c} in this work).
        \newpar

        This resolves the source of another apparent major difference between the two radio classes. In the following sections, we present the results from the Monte Carlo simulations of an LGRB world model in which we attempt to synthetically reconstruct the gamma-ray emission properties of the radio-loud and radio-quiet samples of \citetalias{lloyd2019comparison}. We show that the population-property differences between the two classes as enumerated in \S\ref{sec:intro} naturally emerge from the combined effects of the gamma-ray detection threshold, the artificial cuts on the observational data, and the intrinsic correlations between the prompt gamma-ray properties of LGRBs as reported by \citet{shahmoradi2013multivariate, shahmoradi2015short}.

\section{The LGRB World Model}
\label{sec:lgrbWorldModel}

    \begin{table}
    \label{tab:paraPostStat}
    \caption{Pearson's correlation coefficients of the prompt gamma-ray emission properties of LGRBs inferred from the LGRBs cosmic rate estimates of \citet{shahmoradi2019catalog, osborne2020multilevel}, assuming the LGRB rate density models of \citet{hopkins2006normalization} and \citetalias{butler2010cosmic}.\label{tab:paraPostStat}}
    \begin{center}
        \begin{tabular}{c c c}
        \hline
        \hline
        Parameter & \citetalias{hopkins2006normalization}   & \citetalias{butler2010cosmic} \\
        \hline
        \multicolumn{3}{c}{ Redshift Parameters} \\
            \hline
            $z_0$                       & $0.97$             & $0.97$     \\
            $z_1$                       & $4.5$              & $4.00$     \\
            $\gamma_0$                  & $3.4$              & $3.14$     \\
            $\gamma_1$                  & $-0.3$             & $1.36$     \\
            $\gamma_2$                  & $-7.8$             & $-2.92$     \\
            \hline
            \hline
        \multicolumn{3}{c}{Correlation Coefficients} \\
            \hline
            $\rho_{\liso-\epkz}$        & $0.53\pm0.07$     & $0.60\pm0.05$ \\
            $\rho_{\liso-\eiso}$        & $0.93\pm0.01$     & $0.95\pm0.01$ \\
            $\rho_{\liso-\durz}$        & $0.39\pm0.09$     & $0.37\pm0.07$ \\
            $\rho_{\epkz-\eiso}$        & $0.62\pm0.05$     & $0.69\pm0.03$ \\
            $\rho_{\epkz-\durz}$        & $0.29\pm0.04$     & $0.34\pm0.05$ \\
            $\rho_{\eiso-\durz}$        & $0.54\pm0.05$     & $0.50\pm0.05$ \\
            \hline
            \hline
        \end{tabular}
    \end{center}
\end{table}

    \begin{figure*}
        \centering
        \makebox[\textwidth]
        {
            \begin{tabular}{cc}
                \subfloat[]{\includegraphics[width=0.47\textwidth]{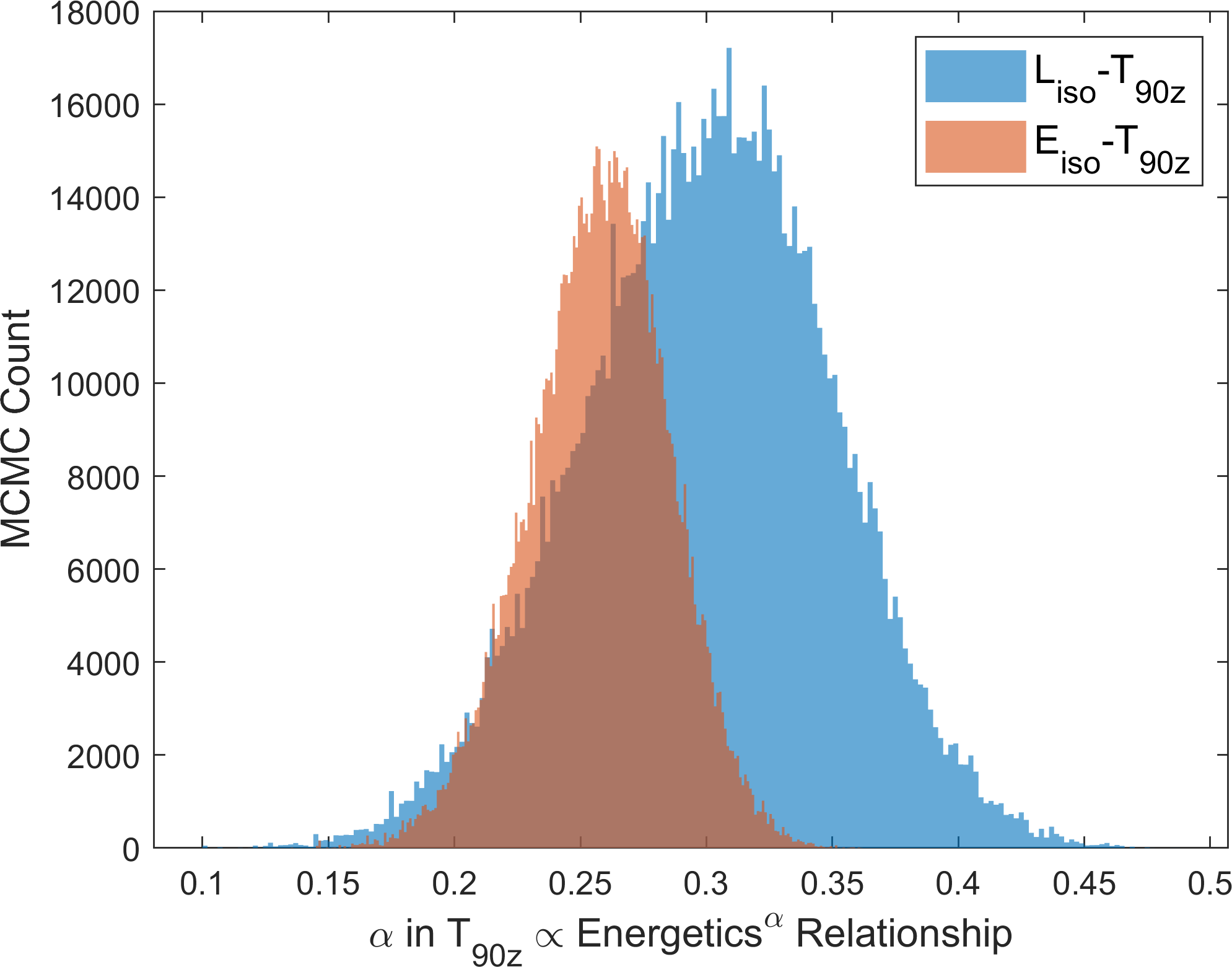}} &
                \subfloat[]{\includegraphics[width=0.47\textwidth]{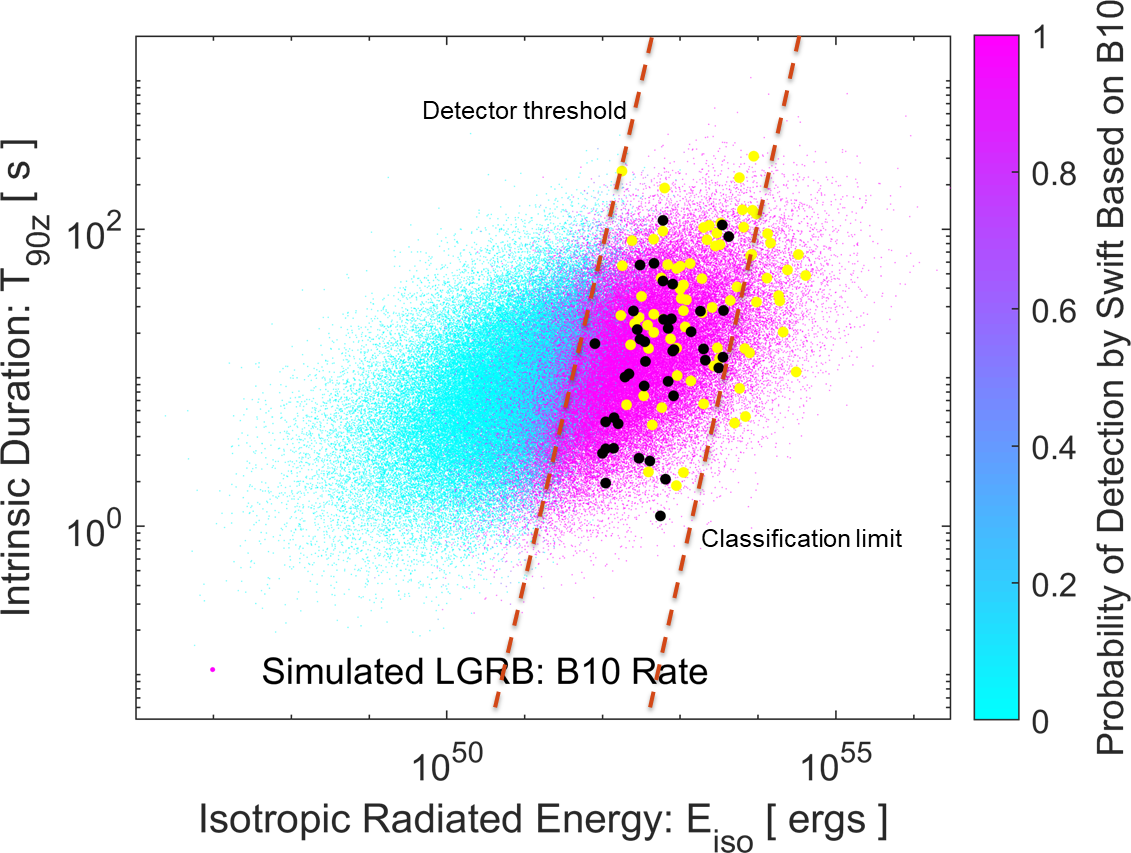}\label{fig:eisodurzs}}  \\
                \subfloat[]{\includegraphics[width=0.47\textwidth]{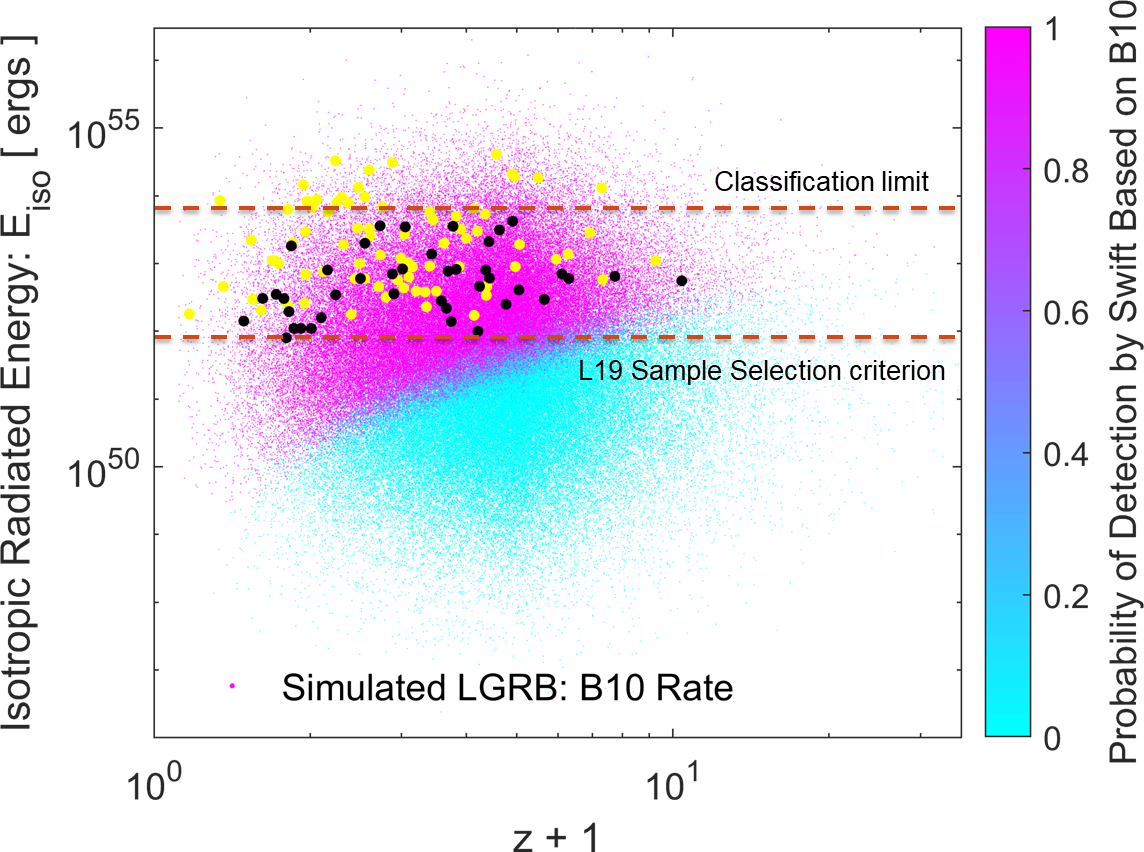}} &
                \subfloat[]{\includegraphics[width=0.47\textwidth]{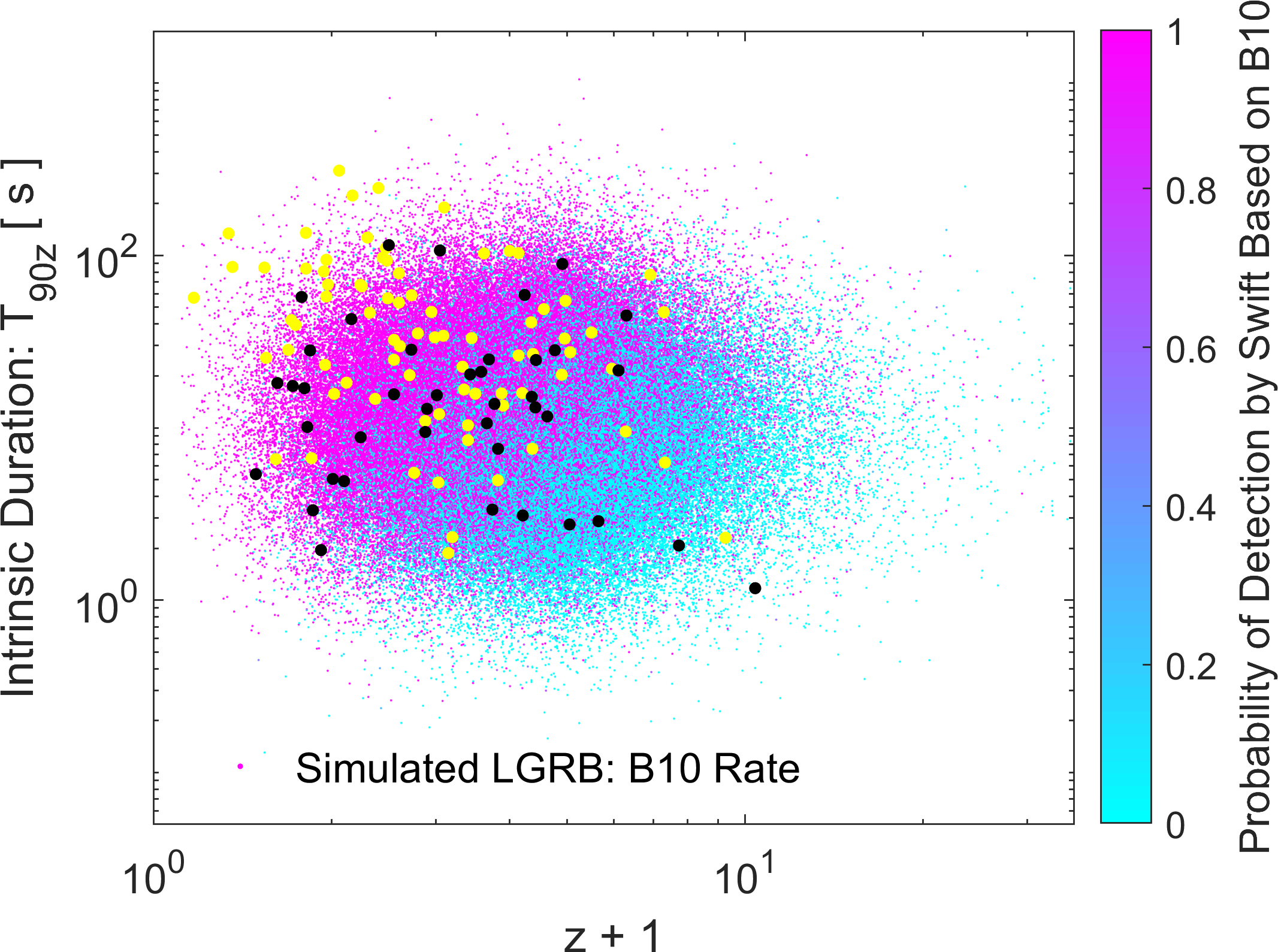}}
            \end{tabular}
        }
        \caption{
        {\bf (a)} The distribution of the exponent of $\liso-\durz$ and $\eiso-\durz$ power-law relationships inferred from modeling the population distribution of 1336 BATSE LGRBs.
        {\bf (b)} An illustration of the predicted underlying intrinsic distribution of LGRBs in the $\eiso-\durz$ plane. Associated with each simulated LGRB is also a probability of it being detectable by the Swift BAT, represented by the cyan-magenta color-map. The overlaid yellow and black points represent the sample of bright radio-loud and radio-quiet LGRBs in \citetalias{lloyd2019comparison} collected such that $\eiso\gtrsim10^{52}$ [ergs]. The effective detection threshold of Swift BAT and the classification limit that separates the two classes of radio-loud and radio-quiet LGRBs are shown by the brown dashed lines. These two limits together potentially shape the narrow distribution of radio-quiet sample (black dots on this plane).
        {\bf (c)} An illustration of the predicted underlying intrinsic distribution of LGRBs in the redshift $(z+1)-\eiso$ plane. The color codings of the plot objects are the same as those of plot (b). The two dashed lines represent the effective limits that are potentially shaping the distribution of radio-quiet sample of radio-quiet LGRBs in this plot.
        {\bf (c)} An illustration of the predicted underlying intrinsic distribution of LGRBs in the $(z+1)-\durz$ plane. The color codings of the plot objects are the same as those of plots (b) and (c).
        \label{fig:monteCarloUniverse}
        }
    \end{figure*}

    In a series of works, \citet{shahmoradi2013multivariate, 2013arXiv1308.1097S, shahmoradi2015short, shahmoradi2019catalog, 2019arXiv190306989S, osborne2020multilevel, 2020arXiv200601157O} have presented evidence for the existence of a significant positive $\eiso-\durz$ correlation of strength $\rho\sim0.5-0.6$ in both populations LGRBs and SGRBs. Here, we build upon these works to create a Monte Carlo universe of LGRBs which are subjected to the detection thresholds of BATSE and Swift Burst-Alert-Telescope (BAT) detector thresholds. Since the majority of the events in the radio-loud and radio-quiet LGRB samples of \citetalias{lloyd2019comparison} belong to the Swift catalog, here we only present the results for the case of the Swift BAT.
    \newpar

    \citet{shahmoradi2013multivariate, shahmoradi2015short} model the joint population distributions of 1366 BATSE catalog LGRBs in the 5-dimensional LGRB property space of,

    \begin{enumerate}
        \item redshift ($z$),
        \item the intrinsic bolometric 1-second isotropic peak energy luminosity ($\liso$) and its equivalent observable, the peak energy flux ($\pbol$).
        \item the intrinsic total isotropic bolometric emission ($\eiso$) and its equivalent observable, the bolometric fluence ($\sbol$).
        \item the intrinsic spectral peak energy ($\epkz$) and its equivalent observable, the observed spectral peak energy ($\epk$).
        \item the intrinsic prompt gamma-ray duration as measured by the time interval during which $90\%$ of the total gamma-ray energy of the LGRB is released ($\durz$) and its equivalent observable, the observed duration ($\dur$).
    \end{enumerate}

    while carefully taking into account the intrinsic correlations between the LGRB prompt gamma-ray properties and the detection threshold of BATSE Large Area Detectors (LADs).
    \newpar

    To create our Monte Carlo universe of LGRBs, we use the inferred posterior distribution of the parameters of their multivariate model under the hypothesis of LGRBs following the LGRB rate density $\mz$ inferred by \cite{butler2010cosmic},

     \begin{equation}
        \label{eq:mz}
        \mz(z) \propto
        \begin{cases}
            (1+z)^{\gamma_0} & z<z_0 \\
            (1+z)^{\gamma_1} & z_0<z<z_1 \\
            (1+z)^{\gamma_2} & z>z_1 ~, \\
        \end{cases}
    \end{equation}

    \noindent where the parameters, ($z_0$,$z_1$,$\gamma_0$,$\gamma_1$,$\gamma_2$) for this equation are (0.97,4.00,3.14,1.36,-2.92). The rate density estimate of \cite{butler2010cosmic} is based on a careful multivariate modeling of Swift catalog of LGRBs. Previously we have used five other rate density models in \cite{osborne2020multilevel}, but found that the rate density estimate of \citetalias{butler2010cosmic} resulted in more accurate results than other models in predicting the redshifts of the BATSE catalog of LGRBs. Table \ref{tab:paraPostStat} summarizes the inferred Pearson's correlation coefficients between the four main prompt gamma-ray attributes of LGRBs considered in our LGRB world model. We refer the interested reader to \citet{shahmoradi2013multivariate, shahmoradi2015short, shahmoradi2019catalog, 2019arXiv190306989S, osborne2020multilevel, 2020arXiv200601157O} for a comprehensive discussion of the modeling approach, and to \citet{2020arXiv200914229S, 2020arXiv201000724S, 2020arXiv201004190S} for details of the MCMC sampling techniques used to construct the parameters posterior distribution.
    \newpar

    Once we have a constrained parametric model for the joint distribution of the population distribution of LGRBs, we generate a Monte Carlo universe of LGRBs by randomly selecting a set of parameters for the LGRB world model from the posterior distribution of parameters and then generating a set of LGRB attributes ($\liso$, $\eiso$, $\epkz$, $\durz$, $z$) given the randomly-selected set of parameters for the LGRB world model.
    \newpar

    Plot (a) of Figure \ref{fig:monteCarloUniverse} displays the distribution of the inferred exponent $\alpha$ for the power-law relationship between the intrinsic duration ($\durz$) and energetics of LGRBs as measured by $\liso$ and $\eiso$. A realization of the $\eiso-\durz$ relation is also displayed in plot (b) of Figure \ref{fig:monteCarloUniverse}.

    \subsection{The $\eiso-\durz$ correlation in radio-loud and radio-quiet LGRBs}
    \label{sec:methods:eisodurzcorr}

        Similar to \citetalias{lloyd2019comparison}, we confirm the existence of a weaker $\eiso-\durz$ relationship in the population of radio-loud LGRBs ($\rho\sim0.23\pm0.11$) compared to the radio-quiet sample of LGRBs ($\rho\sim0.45\pm0.12$). However, bootstrapping results as depicted in plot (g) of Figure \ref{fig:bootL19} indicate that the difference in the correlation strengths between the two radio classes is insignificant. Indeed, there is $10\%$ probability that the underlying $\eiso-\durz$ correlation in the radio-loud sample could be stronger than the corresponding correlation in the radio-quiet sample.
        \newpar

        We have already shown in the previous sections of this manuscript and in \citet{shahmoradi2013multivariate, shahmoradi2015short, shahmoradi2019catalog, 2019arXiv190306989S, osborne2020multilevel, 2020arXiv200601157O} that there is likely a strong intrinsic $\eiso-\durz$ correlation in both LGRB and SGRB classes. This prediction readily explains the existence of a positive $\eiso-\durz$ correlation in both samples of radio-loud and radio-quiet LGRBs. The strengths of the observed correlations are, however, much weaker than the predictions of our LGRB world model because of the strong effects of sample-incompleteness on the radio loud and radio-quiet LGRB samples.
        \newpar

        Furthermore, the slight increase in $\eiso-\durz$ correlation strength in the radio-quiet sample relative to the radio-loud sample of LGRBs can be also potentially explained away in terms of the subtle effects of radio classification and sample-incompleteness on the two LGRB populations. These two artificial fuzzy cuts on the radio-quiet sample are schematically illustrated by the brown dashed lines in plot (b) of Figure \ref{fig:monteCarloUniverse}. The $\eiso$ distribution of radio-quiet LGRBs is likely more affected by the detection threshold of Swift because radio-quiet LGRBs are generally less energetic.
        \newpar

        However, unlike he radio-loud sample, the radio-quiet LGRBs are also limited by an upper $\eiso$ fuzzy threshold which is purely due to the classification of LGRBs into two classes of radio-loud and radio-quiet. If an LGRB is bright enough in gamma-ray, its radio-emission will also become bright enough to be detectable. Therefore, gamma-ray-bright LGRBs are automatically classified as radio-loud LGRBs. The two aforementioned artificial cuts on the radio-quiet sample create a narrow distribution of LGRBs in the $\eiso-\durz$ which artificially increases the observed $\eiso-\durz$ correlation strength of radio-quiet LGRBs compared to the radio-loud sample.

    \subsection{The $z-\eiso$ correlation in radio-loud and radio-quiet LGRBs}
    \label{sec:methods:zeisocorr}

        We confirm the existence of a weaker $z-\eiso$ correlation in the population of radio-loud LGRBs ($\rho\sim0.08\pm0.12$) compared to the radio-quiet sample of LGRBs ($\rho\sim0.33\pm0.13$). However, bootstrapping results as depicted in plot (h) of Figure \ref{fig:bootL19} indicate that the difference in the correlation strengths between the two radio classes is insignificant. Indeed, there is $9\%$ probability that the underlying $\eiso-\durz$ correlation in the radio-loud sample could be stronger than the corresponding correlation in the radio-quiet sample.
        \newpar

        Furthermore, we hypothesize that the slight increase in $z-\eiso$ correlation strength in the radio-quiet sample relative to the radio-loud sample of LGRBs can be also again potentially explained away in terms of the subtle effects of classification and sample-incompleteness on the two LGRB populations. These two artificial fuzzy cuts on the radio-quiet sample are schematically illustrated by the brown dashed lines in Plot (c) of Figure \ref{fig:monteCarloUniverse}. The $\eiso$ distribution of radio-quiet LGRBs is likely more affected by the detection threshold of Swift because radio-quiet LGRBs are generally less energetic.
        \newpar

    \subsection{The $z-\durz$ correlation in radio-loud and radio-quiet LGRBs}
    \label{sec:methods:zdurzcorr}

        Similar to \citetalias{lloyd2019comparison}, we confirm the existence of a stronger $(z+1)-\durz$ anti-correlation in the population of radio-loud LGRBs ($\rho\sim-0.35\pm0.10$) compared to the radio-quiet sample of LGRBs ($\rho\sim-0.07\pm0.18$). However, bootstrapping results as depicted in plot (i) of Figure \ref{fig:bootL19} indicate that the difference in the correlation strengths between the two radio classes is insignificant. Indeed, there is $9\%$ probability that the underlying $z-\durz$ correlation in the radio-loud sample could be weaker than the corresponding correlation in the radio-quiet sample.
        \newpar

        Furthermore, the slight increase in $(z+1)-\durz$ anti-correlation strength in the radio-loud sample relative to the radio-quiet sample of LGRBs can be also potentially explained away in terms of the subtle effects of classification and sample-incompleteness on the two LGRB populations combined with the effects of the correlation strengths of $\eiso-\durz$ and $(z+1)-\eiso$ relationships in the two LGRB radio classes.
        \newpar

        To illustrate the effects of $\eiso-\durz$ and $(z+1)-\eiso$ correlations on the $(z+1)-\durz$ correlation, we generate Monte Carlo realizations of the radio-loud and radio quiet LGRB samples, similar to the two observational samples of \citetalias{lloyd2019comparison}. We do so by generating two samples that have the same distributions of redshift and $\eiso$ as those of the observational radio-loud and radio-quiet samples, while fixing their $\eiso-\durz$ and $(z+1)-\eiso$ correlation strengths to their corresponding values in the observational samples. We then leave the $\durz$ distribution of the two synthetic samples to be randomly determined by our Monte Carlo simulations.
        \newpar

        \begin{table*}
    \begin{center}
        \vspace{5mm}
        \caption{Correlations from \citetalias{lloyd2019comparison} and from figure \ref{fig:correlations}. \label{tab:monteCarloSim}}
        \resizebox{\textwidth}{!}{
        \hskip-0.6cm\begin{tabular}{c c c c }
            \hline
            \hline
            work used for correlations & $\eiso-\durz$ correlation & $(z+1)-\eiso$ correlation & $(z+1)-\durz$ correlation \\
            \hline
            \hline
            \citetalias{lloyd2019comparison} Radio loud GRBs & 0.23 & 0.08 & -0.36 \\       
            \citetalias{lloyd2019comparison} Radio quiet GRBs & 0.46 & 0.40 & -0.08 \\      
            This work (correlation around red point) & 0.23 & 0.08 & -0.11 \\               
            This work (correlation around black point) & 0.46 & 0.40 & 0.05 \\             
            \hline
            \hline
        \end{tabular}
        }
    \end{center}
\end{table*}

        The above simulation scheme allows us to isolate the effects of $\eiso-\durz$ and $(z+1)-\eiso$ correlation strengths on the strength of the $z-\durz$ correlation. The results of the Monte Carlo simulations are summarized in Table \ref{tab:monteCarloSim}. As shown in the last column of the table, although the average simulated correlation strength of the $(z+1)-\durz$ relationship for the sample of radio-loud LGRBs does not fully match the corresponding observed value, the simulations indicate that a similar trend in correlation strengths with comparable differences to those of the observational samples can be reproduced purely based on the correlation strengths of $\eiso-\durz$ and $(z+1)-\eiso$ relationships. All of these Monte Carlo simulation results have been obtained without any a prior assumptions on the type of LGRBs, whether radio-loud or radio-quiet.
        \newpar

        In other words, the observed $(z+1)-\durz$ correlation strength difference between the two radio-loud and radio-quiet samples likely has no physical origins but can be largely attributed to a complex combination of multiple sample incompleteness and selection effects in gamma-ray and radio detection, data collection, and redshift measurement.
        \newpar


        \begin{figure}
            \centering
            \includegraphics[width=0.47\textwidth]{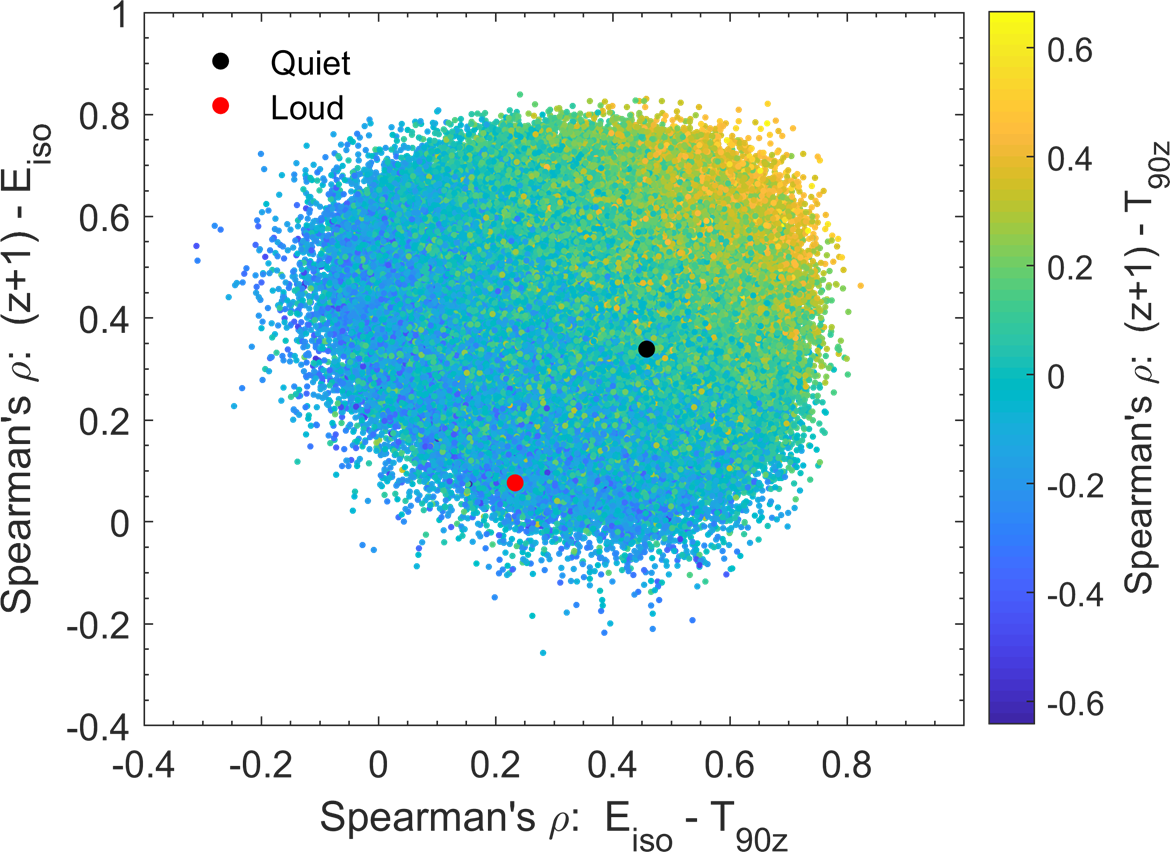}
            \caption{
                An illustration of the Monte Carlo simulations of the correlation coefficient strengths between the three LGRB intrinsic properties: redshift (represented by $z+1$), $\eiso$, and $\durz$. Each point on this plot represents a simulated LGRB sample of comparable size to the radio-loud and radio-quiet sample sizes of \citetalias{lloyd2019comparison}. The two overlaid black and red dots represent respectively, the locations of the radio-loud and radio-quiet samples of \citetalias{lloyd2019comparison} on this plot, with the observed correlation coefficients of $-0.36$ and $-0.08$ for the $(z+1)-\durz$ relationship in the radio-loud and radio-quiet samples, respectively. The observed gradient of correlation strength in this plot is consistent with the observed $(z+1)-\durz$ correlation strengths for the two radio classes.
                \label{fig:correlations}
            }
        \end{figure}

        To better illustrate the the above argument, we depict in Figure \ref{fig:correlations} the distribution of the correlation coefficients between $(z+1)$, $\eiso$, and $\durz$ for all Monte Carlo samples that have simulated. Overlaid on this distribution is the two observed radio-loud and radio-quiet samples of \citetalias{lloyd2019comparison}. From this figure, it is evident that a gradient in the strength of $(z+1)-\durz$ correlation exists as a function of the strengths of the $(z+1)-\eiso$ and $\eiso-\durz$ correlations. We note that this gradient purely results from the intrinsic gamma-ray properties of LGRBs inferred from our LGRB world model. We have made no assumptions on the existence of radio-loud or radio-quiet LGRBs in the aforementioned Monte Carlo simulations.

    \section{Discussion}
\label{sec:discussion}

    The existence of two classes of radio-loud and radio-quiet LGRBs with potentially different progenitors has been recently argued in the literature \citep[e.g.,][]{lloyd2017lack, lloyd2019comparison}. Radio-loud LGRBs have been shown to be on average more energetic, longer-duration and exhibit weaker positive $\eiso-\durz$ but stronger negative $(z+1)-\durz$ correlations than the radio-quiet class of LGRBs (Figure \ref{fig:bootL19}).
    \newpar

    In this work, we have shown that much of the evidence in favor of such radio classification of LGRBs and their distinct progenitors can be purely attributed to the complex effects of detection thresholds of gamma-ray detectors \citep{shahmoradi2009real, shahmoradi2011possible} and radio telescopes \citep{chandra2012radio} on the observed sample of bright LGRBs. Our arguments are built upon the recent discovery of a significant positive $\eiso-\durz$ correlation ($\rho\sim0.5-0.6$) in both populations of LGRBs and SGRBs by \citet{shahmoradi2013multivariate, shahmoradi2015short}. We have shown that the intrinsic $\eiso-\durz$ correlation (Figure \ref{fig:eisodurzs}) along with a potential positive correlation between the gamma-ray and radio luminosity of LGRBs (Figure \ref{fig:chandraSchematic}) are sufficient conditions to generate much of the differing characteristics of radio-loud and radio-quiet LGRBs, without recourse to any radio classification of LGRBs.
    \newpar

    Bootstrapping simulations indicate that some of the proposed spectral and temporal differences between the two proposed radio classes are not statistically significant (\ref{fig:bootL19}). Furthermore, Monte Carlo simulations of the gamma-ray properties of the two radio classes reveal that more than $50\%$ of the reported difference between the $(z+1)-\durz$ correlation strengths of the proposed radio classes can be readily and purely explained in terms of selection effects, sample incompleteness and the strong positive $\eiso-\durz$ correlation in LGRBs.
    \newpar

    In the light of the above arguments, it would seem likely that the presence of the very high energy GeV extended emission in the class of radio-loud LGRBs also results from the overall brighter light-curves of such LGRBs across all energy wavelengths, from radio to GeV.
    \newpar

    The question of whether the radio telescopes have been sensitive enough to detect the faint LGRB radio emissions has been raised previously \citep[e.g.,][]{chandra2012radio, resmi2017radio}. Given the proximity of the $3\sigma$ radio non-detection limits to the observed sample (Figure \ref{fig:ChandraRadioDetectionEfficiency}), it is conceivable that future radio telescopes with increased sensitivities will be able to detect the radio afterglows of more LGRB events \citep{chandra2016gamma}. The numerical simulations of \cite{burlon2015ska} also appear to support this conclusion. The future projects such as the Square Kilometer Array \citep{carilli2004science, johnston2008science} (whose operation is expected in 2027) and the recent upgrades to the existing telescopes such as the Giant Metrewave Radio Telescope \citep{swarup1991giant, ananthakrishnan1995giant, gupta2014gmrt} and many others \citep[e.g.,][]{gupta2017upgraded} will provide a definite answer to the problem of radio-classification of LGRBs.

    \section*{Acknowledgements}

        We thank Nicole Lloyd-Ronning at The University of New Mexico and Poonam Chandra at Tata Institute of Fundamental Research for kindly making their observational GRB data available to us. This work would have not been accomplished without the vast time and effort spent by many scientists and engineers who designed, built and launched the gamma-ray and radio observatories and were involved in the collection and analysis of GRB data.\newpar

    \bibliographystyle{mnras}
    \bibliography{../../../libtex/all}

    \appendix

\end{document}